\documentclass[prd,twocolumn,showpacs,nofootinbib]{revtex4-1}

\usepackage{dcolumn}
\usepackage{amssymb,amsmath}
\usepackage[x11names,table]{xcolor}
\usepackage{float}
\usepackage{natbib}
\usepackage{booktabs}
\usepackage{times}
\usepackage{graphicx}
\usepackage[usenames,dvipsnames]{pstricks}
\usepackage{braket}
\usepackage{color}
\usepackage{textcase}
\usepackage[colorlinks=true,linkcolor=blue,citecolor=blue]{hyperref}
\graphicspath{{Img/}} 
\usepackage{subfig}

\definecolor{airforceblue}{rgb}{0.36, 0.54, 0.66}
\definecolor{royalblue(traditional)}{rgb}{0.0, 0.14, 0.4}
\definecolor{richcarmine}{rgb}{0.84, 0.0, 0.25}
\definecolor{tomato}{rgb}{1.0, 0.39, 0.28}
\definecolor{teal}{rgb}{0.0, 0.5, 0.5}
\definecolor{smalt(darkpowderblue)}{rgb}{0.0, 0.2, 0.6}
\definecolor{amber(sae/ece)}{rgb}{1.0, 0.49, 0.0}
\definecolor{applegreen}{rgb}{0.55, 0.71, 0.0}
\definecolor{brightgreen}{rgb}{0.4, 1.0, 0.0}
\definecolor{brightmaroon}{rgb}{0.76, 0.13, 0.28}
\definecolor{coral}{rgb}{1.0, 0.5, 0.31}
\definecolor{brickred}{rgb}{0.8, 0.25, 0.33}
\definecolor{brightcerulean}{rgb}{0.11, 0.67, 0.84}
\definecolor{cerulean}{rgb}{0.0, 0.48, 0.65}
\definecolor{cyan(process)}{rgb}{0.0, 0.72, 0.92}

\newcommand{\lp}{\left(}
\newcommand{\rp}{\right)}
\newcommand{\lc}{\left[}
\newcommand{\rc}{\right]}
\newcommand{\lk}{\left\{}
\newcommand{\rk}{\right\}}

\newcommand{\be}{\begin{equation}}
\newcommand{\ee}{\end{equation}}
\newcommand{\bse}{\begin{subequations}}
\newcommand{\ese}{\end{subequations}}
\newcommand{\bary}{\begin{eqnarray}}
\newcommand{\eary}{\end{eqnarray}}
\newcommand{\beq}{\begin{equation*}}
\newcommand{\eeq}{\end{equation*}}

\begin{document}
\preprint{APS}
\title{Differentiating short Gamma-Ray Bursts progenitors through multi-MeV neutrinos}

\author{G. Morales } \altaffiliation{E-mail:
gmorales@astro.unam.mx}
\author{N. Fraija }\altaffiliation{E-mail:
nifraija@astro.unam.mx}
\affiliation{Instituto de Astronom\' ia, Universidad Nacional Aut\'onoma de M\'exico, Circuito Exterior, C.U., A. Postal 70-264, 04510, Ciudad de M\'exico, M\'exico}

\date{\today}

\begin{abstract}
With the most recent multi-messenger detection, a new branch in modern astronomy has arisen. The GW170817 event, together with the short gamma-ray burst GRB 170817A, was the first-ever detection of the gravitational waves and its electromagnetic counterpart. These detections encourage us to think that in the following years, we will detect a single event through three different channels: including the mentioned above plus neutrinos from multiple astrophysical sources, like those detected from SN1987A. It is believed that short GRBs are originated in the merger of a black hole with a neutron star (BH-NS) or a neutron star-neutron star (NS-NS) scenario.  Particularly only in the latter case, several simulations suggest that the magnetic field can be amplified up to $\sim 10^{15} - 10^{16}$ G.   Considering this effect over created thermal neutrinos during the initial stage, we study the means of discriminating among short GRB progenitors through the neutrino expected flavor ratio and the opacity created by the
baryon-loaded winds ejected in each scenario. We found that it is more feasible to detect neutrinos from BH-NS mergers than those originated from NS-NS. For instance,  20 MeV-neutrinos created during an NS-NS merger are not released when they propagate with half-opening angles greater than $\sim62^\circ$ concerning the jet launch axis. Finally, we also estimate the number of neutrino events expected on ground-based detectors, finding that it is possible to detect neutrinos from energetic sources  $(L\gtrsim 10^{52}$ erg s$^{-1})$ located close enough, such as the GRB 170817A event, with the future Hyper-Kamiokande detector.
\pacs{XXX}
\date{\today}

\end{abstract}

\maketitle


\section{Introduction}
The detection of a  neutrino burst  in the $(1$--$30)$ MeV range associated with supernova SN1987A \citep{SN1987A} constitutes the beginning of the new  multi-messenger (photons and neutrinos) observation era, and with the first detection of gravitational waves (GW) in 2015 by the Advanced LIGO Collaboration (aLIGO), a second boom gives rise in the multi-messenger scenario \citep{GW150914}.  During the third running (O3) of aLIGO plus Advanced Virgo, a GW signal coming from a binary NS merger was detected on 2017 August 17 at 40 Mpc in the nearby galaxy NGC 4993 \citep{abott17}; shortly after, this signal was associated with an electromagnetic counterpart from the low-luminosity GRB 170817A, and a multi-wavelength follow-up campaign started. Nevertheless, no neutrino signal was detected in temporary or spatial coincidence from this source in the energy range of GeV - EeV \citep{alb17}. \\

In that context, gamma-ray bursts (GRBs) are undoubtedly one of the most exciting astrophysical transients and neutrino sources, coming to be the most luminous explosion in the Universe. In these events, an enormous amount of energy (up to isotropy-equivalent $10^{55}$ erg \citep{att17}) is released during a short timescale (from milliseconds to minutes). Since their discovery in the late 1960s \citep{kle73}, astronomers have spent many resources studying GRBs' phenomenology, which led to the conclusion that these events are from an extragalactic origin and  present a bimodal distribution on their duration \citep{berger2014short}. In this way, GRBs are classified into two subcategories: long and short ones  \cite[e.g., see][for reviews]{2004IJMPA..19.2385Z, 2015PhR...561....1K}. Long gamma-ray bursts (lGRBs) are those events that last over two seconds, while short gamma-ray bursts (sGRBs) correspond to those with a duration typically less than two seconds.  It is widely accepted that progenitors that give rise to both kinds of GRBs are different. In the first case, it is believed that lGRBs are mostly associated with the collapse of a massive star (collapsar model) \citep{hjo03, woo06, hjo12}, whereas sGRBs are produced due to the merger of a binary compact object system either in both BH--NS or NS--NS configuration \citep{eichler1989nucleosynthesis,lee04,lee05,lee2007progenitors,nak07}. This given fact leads to some differences, for instance, numerical simulations suggest that , on one hand, mergers with a larger mass ratio as the case of BH--NS ($E_{\rm iso}\gtrsim 10^{51}\,{\rm erg \ s}^{-1}$)  produces a brighter electromagnetic counterpart and a systematically smaller effective kick \citep{zra13} and on the other hand, during the coalescence of a binary NS system, the magnetic field can be amplified up to several orders of magnitude, reaching values as high as $10^{16}$ G  via Kevin-Helmholtz instabilities and turbulent amplification \citep{Price719,gia09,obe10,kiu14,kiu15}.\\ 

However, so far it is no clear how we could differentiate between the initial binary configuration through direct observations. This is mainly due to the fact that the progenitor remains hidden during the initial phase, because of the great photon opacity. In this sense, it is important to study these bursts through different channels, for instance, neutrinos. Thus, in this work, we propose a mechanism to achieve this goal by studying multi-MeV neutrinos generated in these events. We focus on how their properties get modified when they propagate in media with different physical conditions.  Furthermore, we study the effect of magnetic field amplification in the progenitor's neutrino opacity for the same purpose. Finally, we present the number of expected MeV-neutrino events with current and future ground-based detectors. It is important to mention that hereafter, we use the convention $Q_{\rm x}\equiv Q/10^{\rm x}$  in c.g.s. units, as well as, we adopt the natural units system where, speed of light, reduced Planck constant, and Boltzmann constant are all equal to the unity, $c=\hbar=k_B=1$. As a summary,  in Section \ref{sec:GRB}, we give an overview of sGRBs as well as their main characteristics, while in Section \ref{sec:Neutrino}, we present the neutrino properties split up into: i)  effective neutrino potential, ii) neutrino oscillation theory, and iii) neutrino production processes. The mechanisms involved during the differentiation of sGRBs progenitors are presented and discussed in Section \ref{sec:Diff}. In Section \ref{sec:detection}, we introduce the neutrino detection theory, as well as the neutrino detectors considered in this work. Finally, we discuss and  present our conclusions in Section \ref{sec:conclusion}.

\section{Short Gamma Ray Bursts}\label{sec:GRB}
With a typical time variability from milliseconds up to a couple of seconds and according to the observed energy released during such events, sGRBs are the product of the merger of two compact sources. Because of energy losses by gravitational-wave emission, GRBs' most popular progenitor model is the merger of compact objects; BH-NS or binary NS merger.  In the binary NS merger, the expected remnant is a BH and an accreting disk, although in some cases \citep[NSs with $\sim 2\,\, M_\odot$;][]{2010Natur.467.1081D, 2013Sci...340..448A}, a transitory or stable highly magnetized NS could be formed \citep{1992ApJ...392L...9D,2008MNRAS.385.1455M}. In the case of the BH-NS merger, a  BH with a surrounding accreting disk could appear when the NS is tidally disrupted out of the BH's horizon.  High-accretion rate and rapid angular momentum play a relevant role in the energy extraction via $\nu\bar{\nu}$ annihilation \citep{lee2007progenitors} or MHD processes \citep{1977MNRAS.179..433B}, and the formation/collimation of a relativistic jet.\\

Based exclusively on a binary NS merger, two kinds of GRBs are generally discussed;  low-luminosity (llGRBs) and typical GRBs \citep{2017ApJ...848L..34M}. On the one hand, llGRBs are produced by a mildly relativistic outflow due to the jet is hampered in the advancement by the wind expelled from the hypermassive neutron star (HMNS),  giving rise to a low-luminosity sGRB with $E_{\rm iso}\simeq 10^{46}$ - $10^{47}$ erg \citep{2003MNRAS.343L..36R, 2002MNRAS.336L...7R, 2014ApJ...784L..28N}. On the other hand, typical GRBs are generated by a relativistic jet whose initial conditions are not altered because the collapse to a black hole occurs fast. In this case, the isotropic energy is expected in the range of $E_{\rm iso}\simeq 10^{50}$ - $10^{51}$ erg \citep{2005ApJ...625L..91R,2014ApJ...788L...8M, 2016ApJ...831...22F}. Several studies have been performed to describe the main accretion models in sGRBs \citep{pop99,nar01}. According to them, during the initial stage, a considerable amount of the accretion energy will be converted into neutrinos within the rotation axis's vicinity. Subsequently, they will enhance some fraction of their acquired energy in low mass density regions, mainly by neutrino annihilation processes, but due to neutrinos have a small cross-section, they cannot transfer linear momentum to the baryons contained in the fireball, leading to the creation of an outflow of neutrino-driven baryon-loaded winds that expand anisotropically during the compact-object merger \citep{rosIII,des08}. The mass rate loss during these events is of the order of $10^{-3}-10^{-4}\ M_\odot \rm\ s^{-1} \rm sr ^{-1}$ within the first few milliseconds, being this effect non-negligible during the evolution of the system. In a BNS merger scenario, a strong magnetic field's presence increases the effect on the neutrino-driven winds outflow considerably, and global HD and MHD simulations, such as \citep{perego2014neutrino, siegel2014magnetically},  must be performed in order to know the main properties of these winds in both scenarios.\\

It is widely accepted that GRBs empower a sharp-shaped conical relativistic jet, and based on its line of sight concerning the observer/detector, GRBs can also be classified into \textit{off-axis} and \textit{on-axis} GRBs. For instance, observers whose field of view is within jet opening-angle ($\theta_{\rm obs}<\theta_j$)  see the same burst, but beyond $\theta_j$, the jet emission decrease abruptly, while both prompt emission and the afterglow are very weak as well. Moreover, because of this effect, certain energetic GRBs viewed off-axis could be equipable with some faint llGRBs viewed on-axis, such as GRB 980425,  GRB031213, and GRB170817A, in which case it is necessary to correct this effect using a proper relativistic transformation in the local rest frame \citep{gra05,ram05}. It is worth noting that, in principle, a low-luminosity sGRB viewed on-axis would seem to be a typical sGRB viewed off-axis \citep{2019ApJ...871..123F, 2019ApJ...871..200F, 2019arXiv190407732F}. \\

In order to incorporate GRB dynamics in our study, we rely on the most widely accepted model called ``fireball'' \citep{cav78}, which represents the connection between the relativistic energy outflow and the GRB central engine. This model requires the liberation of a significant radiation concentration in a small volume of practically baryon-free space \citep{PIRAN1999575}. \\


In the sGRBs framework,  after the initial merger is completed, a spinning disk of debris, from a few to dozens of solar masses,  might be left around the BH.  Because the temperature is larger than the $e^\pm$ pair production, nuclei are photo-disintegrated, and the plasma consists mainly of free $e^\pm$ pairs, $\gamma$-ray photons, and baryons \citep{2004ApJ...608L...5L}.  The so-called fireball plasma connected to the progenitor is formed as the base of the jet.
According to the fireball model,  two stages are expected; the prompt emission: when inhomogeneities in the jet lead to internal collisionless shocks \citep[when material launched with low velocity is cached by material with high-velocity;][]{1994ApJ...430L..93R,2017ApJ...848...15F}  and the afterglow: when the relativistic outflow sweeps up enough external material \citep{1997ApJ...476..232M,2015ApJ...804..105F,2016ApJ...818..190F,2017ApJ...848...94F,2019ApJ...885...29F, 2019ApJ...883..162F, 2019ApJ...879L..26F}. We want to emphasize that although high-energy neutrinos are producing in the prompt emission and afterglow, in this work, we are only interested in those produced during the merger and the initial fireball by thermal processes in the MeV range. These neutrinos' properties get modified when they propagate in a non-vacuum medium through the \textit{MSW effect} \citep{wol78}, so these additional aftermaths must be taken into account in the study of the neutrino behavior.\\

 Due to plasma ingredients are strongly coupled, the fireball can be considered spherically homogeneous. Initially opaque, this fireball will expand adiabatically by radiation pressure until it becomes transparent to photons and neutrinos. The initial temperature is high enough  ($T\gtrsim1$ MeV) to surpass the binding energy in nuclei, propitiating a medium consisting essentially of free baryons which, among other parameters, such as temperature, electron chemical potential, and magnetic field intensity,  have an important role within the initial neutrino interactions.\\
 
Last but not least, it is crucial to take into account the neutrino opacity within the fireball during the initial phase, which can be described as a function of the optical depth $\tau=r/\lambda$, where $r$ is the fireball radius and $\lambda$ is the mean free path. The mean free path of $\nu_{e,\mu,\tau}$  can be estimated as \citep{2005MNRAS.364..934K} 

 \bary
\lambda_e&=&5.9\times 10^6\, {\rm cm}\,\, T_7^{-5}\,,\cr
\lambda_{\mu,\tau}&=&2.56\times 10^7\, {\rm cm}\,\, T_7^{-5}\,.
\eary
The difference between electron and muon/tau neutrinos corresponds to the fact that $\nu_e$ interacts by charged-current (CC) and neutral current (NC), whereas $\nu_{\mu/\tau}$ only through NC.  Therefore, the optical depths for electron and muon/tau neutrinos are
\bary
\tau_e&=&54\,\, E_{52}^{5/4}\,r_{0,7}^{-11/4}\,,\cr
\tau_{\mu,\tau}&=&7.4\,\, E_{52}^{5/4}\,r_{0,7}^{-11/4}\,.
\eary

\section{Neutrinos}\label{sec:Neutrino}
\subsection{Neutrino effective potential in matter}
As active neutrino propagates in a medium, the dynamics are affected by the effective potentials due to the coherent interactions. These interactions are given by elastic weak charged-current and neutral current  scattering  \citep[see][]{1988NuPhB.307..924N, 1991NuPhB.349..754E}. For a medium immersed in a magnetic field and a heat bath,  the effects are introduced  through Schwinger's proper-time method \citep{1951PhRv...82..664S} and finite-temperature field theory formalism, respectively. The neutrino's effective potential is estimated from its self-energy Feynman diagram.  It is calculated in detail in \cite{2014ApJ...787..140F}.\\
The dispersion relation of neutrino is 
\be\label{diseq}
{\rm V_{eff}=k_0-| \vec{k}|}\,,
\ee
where ${\rm  \vec{k}}$ is calculated through  the neutrino field equation in a medium \citep{1988NuPhB.307..924N}
\be
[ {\rlap /k} -{\Sigma}(k)]\psi_L=0\,.
\ee
The term $\Sigma (k)$ carries the information of medium such as the velocity, the magnetic field and the neutrino momentum.

Following \citep{2014ApJ...787..140F}, the total neutrino self-energy is given by the exchanges of  $W$ boson  (${\rm \Sigma_W(k)}$) and  $Z$ boson ($\Sigma_Z$ and $\Sigma_t$). Therefore, although the total neutrino self-energy can be written as ${\rm \Sigma(k)=\Sigma_W(k)+ \Sigma_Z(k)+\Sigma_t(k)}$, the effective potential useful for neutrino oscillations in a medium is ${\rm \Sigma(k)=\Sigma_W(k)}$ \cite[only due to the CC, $V_{\rm eff}=V_{e}-V_{\mu,\tau}$;][]{2004PhRvD..70d3001B,1998PhRvD..58h5016E, 2009PhRvD..80c3009S,  2009JCAP...11..024S, wol78, 1992PhRvD..46.1172D}. In other words,  the effective potential will be only dependent on electron density. \\  
The neutrino self-energy of the  W-boson exchange is \citep{2014ApJ...787..140F}
\be
-i\Sigma(k)={\mathcal R}\,\biggl[\frac{g^2}{2}\int\frac{d^4 p}{(2\pi)^4}\, \gamma_\mu S_{\ell}(p)\gamma_\nu\,
 \,W^{\mu \nu}(q)\biggr]\,{\mathcal L}\,,
\label{Wexch}
\ee
where $g^2=4\sqrt2 G_Fm_W^2$  is the weak coupling constant with $m_W$ the W-boson mass and G$_F$ the Fermi coupling constant, $W^{\mu\nu}$ is the W-boson propagator in unitary gauge \citep{1998PhRvD..58h5016E,  2009JCAP...11..024S},  $S_l(p)$  is the charged lepton propagator \citep{2014ApJ...787..140F}.  It is worth noting that although neutrino does not have a charge, the magnetic field interacts with the charged particles in a medium, and the propagator of charged lepton carries this information. The terms $\gamma_\mu$ are the Dirac's matrices, and ${\mathcal R}$ and $ {\mathcal L}$ are the right and left projection operators, respectively.\\
Calculating the real part of neutrino self-energy  (Equation \ref{Wexch})  ${\rm Re \Sigma(k)= {\mathcal R}\,[a_{\perp} \rlap /k_\perp + b \rlap /u + c \rlap /b]\, {\mathcal L}}$ as a function of the Lorentz scalars ($a_{\perp}$, $b$ and $c$), the dispersion relation (Equation \ref{diseq}) is in the form 
\be
{\rm V_{eff}=b-c\,\cos\varphi-a_{\perp}|{\vec {k}}|\sin^2\varphi},
\label{poteff}
\ee
where $\varphi$ is the angle between the neutrino momentum and the direction of magnetic field.  The Lorentz scalars for  the strong and weak magnetic field limit  are computing in the appendix.
\subsubsection{Strong $\vec{B}$ limit}
From Equation \ref{poteff}, the neutrino effective potential in the strong magnetic field regime becomes
\begin{eqnarray}\label{eq:veffs}
V_{\rm eff,s}=\frac{\sqrt2\,G_F\,m_e^3 B}{\pi^2\,B_c}\biggr[\sum^{\infty}_{l=0}(-1)^l\sinh\alpha_l   \left[F_s-G_s\cos\varphi \right]\nonumber\\
-4\frac{m^2_e}{m^2_W}\,\frac{E_\nu}{m_e}\sum^\infty_{l=0}(-1)^l\cosh\alpha_l  \left[J_s-H_s\cos\varphi \right]  \biggr]\,,\cr
 \end{eqnarray}
where $m_e$ is the electron mass, $\alpha_l=(l+1)\mu/T$  with $\mu$ and $T$ the chemical potential and temperature, respectively, $B_c=m_e^2/e=4.141\times 10^{14}\, {\rm G}$ is the critical magnetic field,  $E_\nu$ is the neutrino energy and the functions F$_s$, G$_s$, J$_s$, H$_s$ are in appendix.
\subsubsection{Weak $\vec{B}$ limit}
Likewise, we found that the neutrino effective potential in the weak magnetic field limit is
\begin{eqnarray}\label{eq:veffw}
V_{\rm eff,w}=\frac{\sqrt2\,G_F\,m_e^3 B}{\pi^2\,B_c}\biggr[\sum^{\infty}_{l=0}(-1)^l\sinh\alpha_l  \left[F_w-G_w\cos\varphi \right]\nonumber\\
-4\frac{m^2_e}{m^2_W}\,\frac{E_\nu}{m_e}\sum^\infty_{l=0}(-1)^l\cosh\alpha_l \left[J_w-H_w\cos\varphi \right]\,.\cr 
\end{eqnarray}
%
%
%
%

\subsection{Neutrino Oscillation}
Neutrino oscillation is a phenomenon widely studied since the second half of the last century. It even nowadays is an active field of research, being the discovery of the massive neutrino properties one of the most important results of modern physics.
In general, this phenomenon refers to a quantum effect in which there is a periodic change between the probability amplitude of an elemental particle created with an eigenstate $\alpha$ and detected with an eigenstate $\beta$, such as $\alpha\neq\beta$. Thus, we are going to describe the oscillations of thermal neutrinos propagating in a fireball medium \citep[where they are produced;][]{2014MNRAS.437.2187F,2015MNRAS.450.2784F}. So, we show a summary of the neutrino oscillation theory in both vacuum and the matter.

\subsubsection{Vacuum}
Neutrinos propagating in the vacuum are not affected by external surrounding particles, and hence their amplitude probability could be easily expressed as \citep{gon03}

\begin{equation}
P(\nu_\alpha\rightarrow\nu_\beta (t)) = \sum_{k>j}U_{\alpha k}^* U_{\beta k}  U_{\alpha j}^*U_{\beta k}\,\,e^{-i(E_k-E_j)t}\,,
\label{vacuum1}
\end{equation}
where $E_k$ is the neutrino dispersion relation, which can be approximated as ${E_k\approx E+({m_k^2}/{2E})}$, with $E=|{\vec{p}}|$ and $E_k-E_j \approx \Delta m_{kj}^2/2E$. In the last expression $\Delta m_{kj}^2$ represents the mass squared differences $\Delta m_{kj}^2\equiv m_k^2-m_j^2$, such that
\begin{equation}
P(\nu_\alpha\rightarrow\nu_\beta (t)) = \sum_{k>j}U_{\alpha k}^* U_{\beta k}  U_{\alpha j}^*U_{\beta k}\,\,e^{-i(\frac{\Delta m_{kj}^2}{2E})t}\,.
\label{vacuum2}
\end{equation}
Likewise, we can assume that  neutrino propagation time is proportional to its distance traveled. Then, the oscillation phase  is determined as $\phi_{kj}=-{\Delta m_{kj}^2L}/{(2E)}\,$ and the oscillation length in the vacuum (typical distance in which the oscillation phase is equal to a 2$\pi$ period) could be expressed as  $L_{\text{osc,v}}=({4\pi E)}/{\Delta m_{kj}^2}$. Thus, in order to have important oscillation effects,  this oscillation length must be greater than the distance between  source and detector; otherwise, we can only  treat average oscillation effects \citep{jarlskog1985c}. 

\subsubsection{Matter}
Wolfenstein demonstrated that neutrinos propagating in a non-vacuum medium are affected by an effective potential, equivalent to the refractive index of that medium  \citep{PhysRevD.17.2369}. Later, Mikheyev and  Smirnov \citep{mikheyev1986yad} showed that the neutrino oscillation parameters are modified when they propagate within a material medium; currently, this is known as \textit{MSW effect}. This additional potential increases the effective neutrino mass, as well as their mass and flavor.

\subsection{Two-neutrino case}

In this case, we consider the neutrino oscillation between eigenstates $\alpha$ and $\beta$ with $\alpha\neq\beta$.  We take into account the equations of oscillation probabilities between electron into muon and tau neutrino ($\nu_e\rightarrow\nu_\mu\,\,$ and $\,\,\nu_e\rightarrow\nu_\tau$).\\ Using the equation of temporal evolution \citep{2014ApJ...787..140F}
\begin{equation}\label{ecevtemp}
i\begin{pmatrix} {\dot{\nu_e}} \\ \dot{\nu_\mu} \end{pmatrix} =\begin{pmatrix} V_{ \text{eff} }-\Delta \text{cos}  2\theta \,   & \, \frac { \Delta  }{ 2 } \text{sin} 2\theta  \\ \frac { \Delta  }{ 2 }\text{sin}  2\theta  & 0 \end{pmatrix} \begin{pmatrix}{\nu_e} \\ {\nu_\mu} \end{pmatrix} \,,
\end{equation}
with \be\Delta= \frac{\Delta m_\nu^2}{2E_{\nu}}=\frac{m_{\nu_e}^2-m_{\nu_\mu}^2}{2E_\nu}\,,\ee and $\theta$ the two-neutrino mixing angle, we get the oscillation probability in a two-neutrino mixing scenario as

\begin{equation}
P_{\nu_e\to\nu_\mu}(t) =|\psi_{\nu_e	\to \nu_\mu}|^2  
=\frac{\Delta^2 \text{sin}^2 2\theta}{\omega^2}\text{sin}^2\left (\frac{\omega t}{2}\right)\, ,\label{prob}
\end{equation}
where $\,\omega=\sqrt{(V_{ \text{eff}}-\Delta \text{cos}  2\theta)^2+(\Delta\text {sin} 2\theta\,)^2},$ and the matter effects are considered using the neutrino effective potential.\\
In this way, the neutrino oscillation length in matter turns out to be 
\begin{equation}\label{l1}
L_{\text{osc,m}}=\frac{L_\text{osc,v}}{\sqrt{\text{cos}^2 2\theta (1-\frac{V_{\text{eff}}}{\Delta \text{cos} 2\theta}
    )^2+\text{sin}^2 2\theta}}\,.
\end{equation} In order to satisfy the resonance condition, we require the positivity of the potential, which implies that $V_{\rm eff}=\Delta\cos 2\theta$, and hence, the resonance length is obtained as $L_{\rm res}=L_{\rm osc, v}/\sin2\theta$.

\subsection{Three-neutrino case}

%
In a three neutrino-mixing scenario, the evolution of  neutrino flavors is governed by the Schr\"odinger equation in which a neutrino state with initial flavor $\alpha$, follows the evolution equation
\be
i\frac{d\vec{\nu}_\alpha}{dt}=H\vec{\nu}_\alpha,
\ee
where the effective Hamiltonian is
\be
H=U\cdot \mathcal{M}^2 \cdot U^\dagger+ \mathcal{A}\,,
\ee
with the matrices given by 
{\small
\begin{equation}
\mathcal{M}^2=\frac{1}{2E_\nu}
\begin{pmatrix}
-\delta m^2_{21} & 0 & 0 \\
 0 &  0 & 0    \\
0&0&\delta m^2_{32} \\
\end{pmatrix}\,,
\end{equation}
}

{\small
\be
\mathcal{A}= 
\begin{pmatrix}

V_{\rm eff} &  0   &  0 \cr
0 &  0  &  0    \cr
0 &  0   &  0\cr
\end{pmatrix}.
\ee
}
The three-neutrino mixing matrix $U$ is given in \citep{gon03}.  The neutrino state is defined as
{\small
\be
\vec{\nu}_\alpha  \equiv
\begin{pmatrix}

\nu_e\cr
\nu_\mu\cr
\nu_\tau\cr
\end{pmatrix},
\ee
}
%
%
%
The amplitude of {\small $\nu_\alpha\to \nu_\beta$} transitions after a time $t$ is {\small$\phi_{\nu_\alpha \to\nu_\beta}(t) =\vec{\nu}_\alpha \vec{\nu}^T_\beta$}  and the probability is given by {\small $P_{\nu_\alpha\to \nu_\beta}(t)=|\phi_{\nu_\alpha \to\nu_\beta}(t)|^2$}, which turns out to be \citep{RevModPhys.75.345}.
 \begin{align}\label{ozzie}
P_{ee}&= 1-4s_{13,m}^2c_{13,m}^2S_{31}\,,\nonumber\\
 P_{\mu\mu}&=  1-4s_{13,m}^2c_{13,m}^2 s_{23}^4 S_{31} -   4s_{13,m}^2s_{23}^2c_{23}^2S_{21}\nonumber\\
&\ \ \ -     4c_{13,m}^2s_{23}^2 c_{23}^2S_{32}\,,\nonumber\\
 P_{\tau\tau}&=  1-4s_{13,m}^2c_{13,m}^2 c_{23}^4S_{31}-   4s_{13,m}^2s_{23}^2c_{23}^2  S_{21}\nonumber\\ &\ \ \  -     4c_{13,m}^2s_{23}^2 c_{23}^2S_{32}   \,,\nonumber\\
 P_{e\mu}&=4s_{13,m}^2c_{13,m}^2 s_{23}^2S_{31} \,,\\
  P_{e\tau}&=4s_{13,m}^2c_{13,m}^2 c_{23}^2S_{31} \,,\nonumber\\
   P_{\mu\tau}&=-4s_{13,m}^2c_{13,m}^2 s_{23}^2c_{23}^2S_{31}+  4s_{13,m}^2s_{23}^2c_{23}^2S_{21}\nonumber\\ &\ \ \   +  4c_{13,m}^2 s_{23}^2c_{23}^2S_{32} \,,\nonumber
    \end{align}
     where $\theta_{13,m}$ is the effecting mixing angle in matter given by
     \be \text{sin}2\theta_{13,m}=\frac{\text{sin}2\theta_{13}}  {{\sqrt[]{\left(\text{cos}2\theta_{13}-\frac{2E_\nu V_{\text{eff}}}{\Delta m_{32}^2} \right)^2+(\text{sin}2\theta_{13})^2}}}\,,
\ee 
and $ S_{ij}$ corresponds to the neutrino oscillation factors defined as \be S_{ij}=\text{sin}^2\Bigg( \frac{\Delta\mu_{ij}^2L}{4E_\nu} \Bigg) \label{3.31}\,.\ee where the term $\Delta \mu_{ij}^2$, represents the squared mass diferences in matters given by the following relations 
\begin{align}
\Delta\mu_{21}^2&=\frac{\Delta m_{32}^2}{2}\Bigg( \frac{\text{sin}2\theta_{13}}{\text{sin}2\theta_{13,m}}-1\Bigg)-E_\nu V_{\text{eff}}\,,\nonumber\\
\Delta\mu_{32}^2&=\frac{\Delta m_{32}^2}{2}\Bigg( \frac{\text{sin}2\theta_{13}}{\text{sin}2\theta_{13,m}}+1\Bigg)+E_\nu V_{\text{eff}}\,,\\
\Delta\mu_{31}^2&=\Delta m_{32}^2\Bigg( \frac{\text{sin}2\theta_{13}}{\text{sin}2\theta_{13,m}}\Bigg)\,.\nonumber
\end{align}

In this case, the  oscillation length of the transition probability  is given by
\be
L_{\rm osc,m}=\frac{L_{\rm osc, v}/\cos^2 2\theta_{13}}{\sqrt{ (1-\frac{V_{\rm eff}}{V_{\rm res} }
    )^2+\tan^2 2\theta_{13}}},
\label{osclength}
\ee
where   $V_{\rm res}=\delta m^2_{32} \cos 2\theta_{13}/2 E_{\nu}$ and $L_{\rm osc, v}=4\pi E_{\nu}/\delta m^2_{32}$ is the vacuum oscillation length. The resonance condition $V_{\rm eff}=  V_{\rm Res}$ can be written as
\be\label{reso3}
V_{\rm eff}= 5\times 10^{-7}\frac{\delta m^2_{32,\rm eV}}{E_{\nu,\rm MeV}}\,\cos2\theta_{13}\,,
\ee
while the resonance length is

\be
L_{\rm res}=\frac{4\pi E_{\nu}}{\delta m^2_{32}\,\sin 2\theta_{13}}\,,
\ee
and the adiabatic condition at the resonance can be expressed as

{\small
\be
\kappa_{\rm res}\equiv  8\pi\, l_{\rm res}^{-2}\, \left (\frac{dV_{\rm eff}}{dr}\right)^{-1} \ge 1\,.
\label{adbcon}
\ee
}

\subsection{Best-fit oscillation parameters }
\subsubsection{{Two-Neutrino mixing}}
Based on appearance and disappearance neutrino oscillation experiments,  fluxes of solar, atmospheric, and accelerator neutrinos have provided values of the squared-mass difference and mixing angles.
\begin{itemize}
\item The best-fit oscillation parameters based on  {\rm  solar experiments} are $\delta m^2=(5.6^{+1.9}_{-1.4})\times 10^{-5}\,{\rm eV^2}$ and $\tan^2\theta=0.427^{+0.033}_{-0.029}$\citep{aha11}.
\item The best-fit oscillation parameters based on  {\rm  atmospheric experiments} are $\delta m^2=(2.1^{+0.9}_{-0.4})\times 10^{-3}\,{\rm eV^2}$ and $\sin^22\theta=1.0^{+0.00}_{-0.07}$\citep{abe11a}.
\item The best-fit oscillation parameters based on  {\rm  accelerator experiments} are \citep{chu02} found two well defined regions of oscillation parameters with either $\delta m^2  \approx  7\, {\rm eV^2}$ or $\delta m^2 < 1\, {\rm eV^2} $ compatible with both LAND and KARMEN experiments, for the complementary confidence and the angle mixing is $\sin^2\theta=0.0049$. In addition, MiniBooNE found evidence of oscillations in the 0.1 to 1.0 eV$^2$, which are consistent with LSND results \citep{1998PhRvL..81.1774, 1996PhRvL..77.3082A}.
\item Combining the solar, atmospheric and accelerator parameters,  the best-fit oscillation parameters are \citep{aha11,wen10} $\delta m_{21}^2= (7.41^{+0.21}_{-0.19})\times 10^{-5}\,{\rm eV^2}$ and $\tan^2\theta_{12}=0.446^{+0.030}_{-0.029}$ for  $\sin^2 \theta_{13} < 0.053$  and  $\delta m_{23}^2=(2.1^{+0.5}_{-0.2})\times 10^{-3}\,{\rm eV^2}$ and $\sin^2\theta_{23}=0.50^{+0.083}_{-0.093}$ for $\sin^2 \theta_{13} < 0.04$.\\
It is worth noting that $\delta m_{21}^2=\delta m_{\rm sol}^2$ and $\delta m_{32}^2=\delta m_{\rm atm}^2$. Here, the solar parameters correspond to the large mixing angle solution.
\end{itemize}
\subsubsection{{Three-Neutrino Mixing}}
We show in Table \ref{table1} a summary of the most current status of neutrino oscillation parameters in a three-flavor mixing scenario performed by global--fit analysis from \citep{salas2018}.

\begin{table}[H]
\centering \rowcolors{1}{}{cyan(process)!18}
\begin{tabular}{cc}\hline\hline\\
 \rowcolor{coral!80}\text{Parameter}                             & \text{Best-fit  $\pm1\sigma$ (NO)}  \\
  \hline\hline  \addlinespace[1mm]\\
$\text{sin}^2\theta_{12}$                      & ${0.320_{-0.016}^{+0.020}}$   \\  \addlinespace[1.2mm]
$\theta_{12}/\ ^\circ$                          & $34.5_{-1.0}^{+1.2}$      \\  \addlinespace[1.2mm]
$\text{sin}^2\theta_{23}$                      & $0.547_{-0.030}^{+0.020}$    \\  \addlinespace[1.2mm]
$\theta_{23}/\ ^\circ$                          & $47.7_{-1.7}^{+1.2}$         \\  \addlinespace[1.2mm]
$\text{sin}^2\theta_{13}$                      & $0.02160_{-0.00069}^{+0.00083}$ \\ \addlinespace[1.2mm]
$\theta_{13}/\ ^\circ$                          & $8.53_{-0.15}^{+0.14}$       \\ \addlinespace[1.2mm]
$\dfrac{\Delta m_{21}^2}{10^{-5}\text{ eV}^2}$ & $7.55_{-0.16}^{+0.20}$       \\ \addlinespace[1.2mm]
$\dfrac{\Delta m_{31}^2}{10^{-3}\text{ eV}^2}$ & $2.50\pm 0.03$ \\ \addlinespace[1.2mm] \hline\hline
\end{tabular}
\caption{Summary of the  neutrino oscillation parameters obtained   from global--fit analysis considering a NO  (Normal Ordering) scheme ($\Delta m_{31}^2>0$).  }\label{table1}
\end{table}

\subsection{Neutrino processes involved}

Due to the high temperatures reached during the initial stage, several neutrino emission within  the plasma fireball takes place, the essential processes are \citep{dic72,lat76}:
\begin{itemize}
\item pairs annihilation ($e^++e^-\to\nu_x+\bar{\nu}_x$),
\item plasmon decay $(\gamma\to\nu_x+\bar{\nu}_x)$,
\item photo-neutrino emission $(\gamma+e^-\to e^-+\nu_x+\bar{\nu}_x)$,
\item positron capture $(n+e^+\to p+\bar{\nu}_e)$,
\item electron capture $(p+e^-\to n+\nu_e)$,
\end{itemize}
 with ($x=e,\mu,\tau$). Given the initial non-zero baryon density, the mean free path of neutrinos in the fireball is constituted principally by the interaction with nucleons and in a non-homogeneous medium could be expressed as a function of the neutrino cross--section  and the baryon density.\\

\begin{figure}[htbp!]
\centering
\includegraphics[width=0.95\columnwidth]{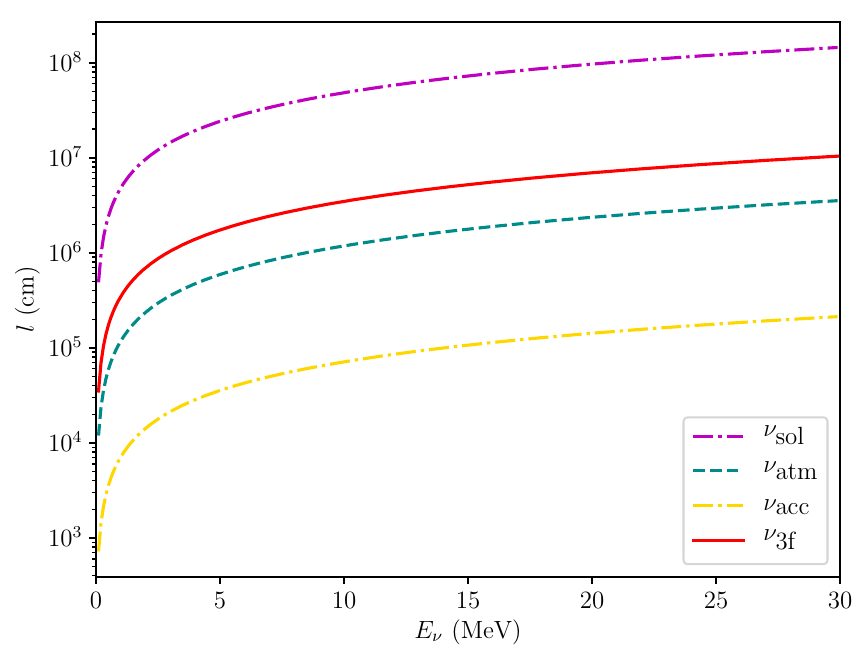}
\caption{Neutrino resonance lengths as a function of $E_\nu$ using, from the top to bottom; solar, atmospheric, accelerator and three-flavor best-fit neutrino parameters shown in Table \ref{table1}.}
\label{plot:RL}
\end{figure}

We show in Figure \ref{plot:RL}, the neutrino resonance lengths as a function of neutrino energy  using  Equations (\ref{l1}) and (\ref{osclength}) in a two and three-neutrino mixing scenario. In these plots, we have used the most recent best-fit neutrino oscillation parameters summarized  in Table \ref{table1}. Excluding the solar neutrino parameters, we find that the resonance lengths for three-flavor neutrinos lie in a range value of $l_{\text{res}}<10^{7}$ cm, which is less than than the typical scale of the fireball size during the initial stage, i.e., these neutrinos will oscillate resonantly before leaving the fireball, implying that they will be released in each kind of flavor almost in the same proportion. \\

\begin{figure*}[htbp!] 
\centering

	\subfloat
					{
  					\includegraphics[width=0.45\textwidth]{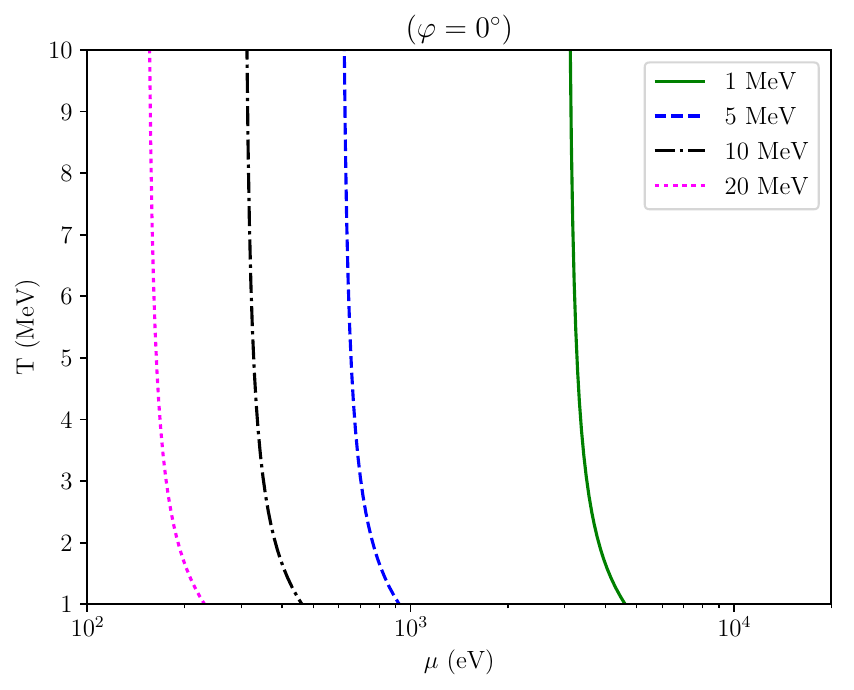}%
          	        }
					\qquad
	\subfloat
					{
  					\includegraphics[width=0.45\textwidth]{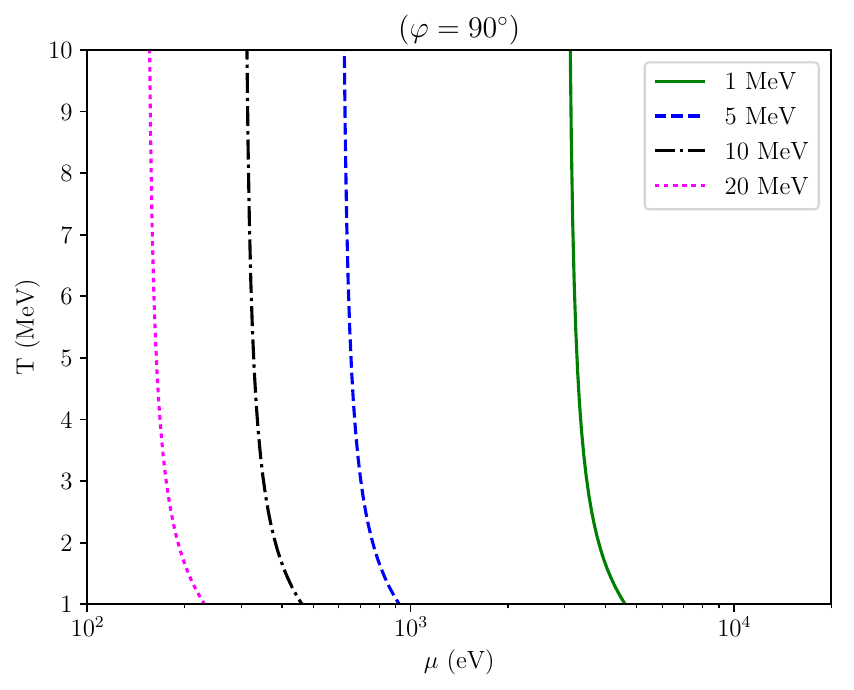}%
          	        } \qquad
	\subfloat
					{
  					\includegraphics[width=0.45\textwidth]{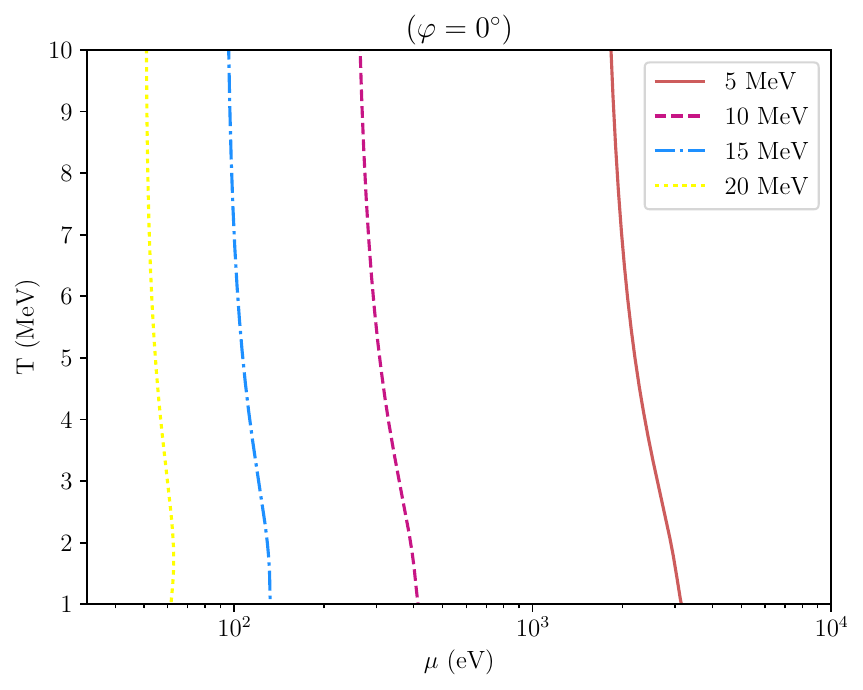}
 				    \label{plot:RCstr_0}
          	        }
					\qquad
	\subfloat
					{
  					\includegraphics[width=0.45\textwidth]{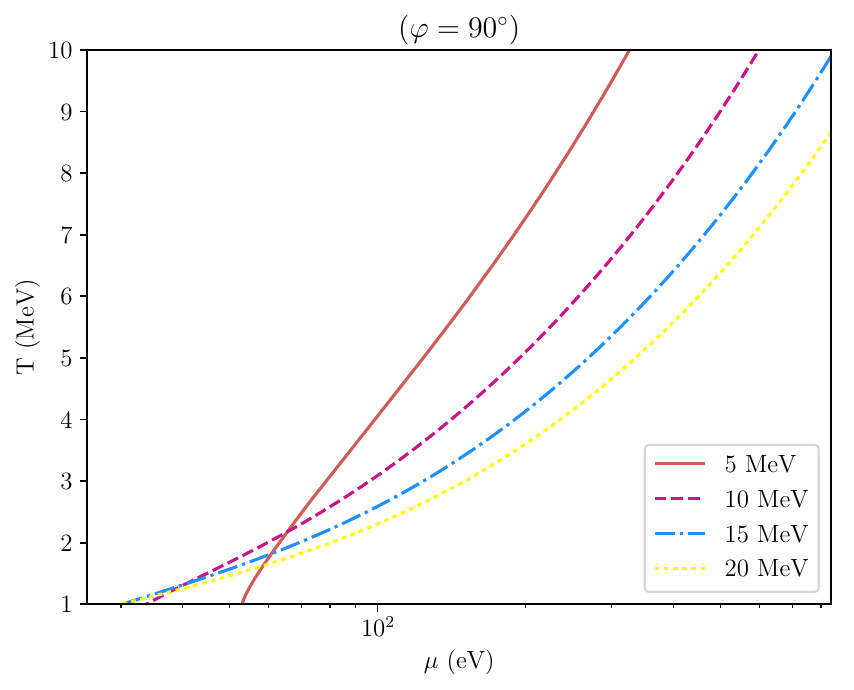}%
 				    \label{plot:RCstr_05}
          	        }    
          	        \qquad  
\caption{Upper panels: Resonance conditions in the BH--NS regime with neutrinos propagating in both parallel (left) and perpendicular (right) direction along the progenitor magnetic field lines for different neutrino energies ($E\nu=\lk1,\ 5,\ 10,\ 15\rk$ MeV). The comparison of these plots shows a non-dependence behavior with the angular neutrino propagation.  \\
Bottom panels: The same from upper panels but for a BNS merger scenario. The magnetic field influence over the neutrino propagation direction modifies the resonance conditions drastically for greater $\varphi$ angles. Considering a strong magnetic limit present in a BNS system, the neutrino effective potential exhibits a sensitive angular dependence regarding the neutrino propagation.}  
\label{plot:RC}        	        
\end{figure*}

These  parameters also allow us to find the precise conditions by which  resonance conditions are presented. In each case, we plot in Figure \ref{plot:RC} these contributions for $B=10^{16}$ G and $B=10^{12}$ G, respectively. We also find that during the magnetic field amplification, the neutrino propagation angle also represents a considerable influence on the potential for greater angles. However, for a BH--NS merger, the contribution of $\varphi$ keeps similar, even for extreme angular values.\\

\section{Differentiating \texorpdfstring{\MakeLowercase{s}}\  GRB\texorpdfstring{\MakeLowercase{s}} \ \  progenitors}\label{sec:Diff}

\subsection{Neutrino oscillations and expected flavors }

Taking the magnetic field amplification into account, we want to know how the neutrino properties are modified when they propagate into a fireball medium. In this context, we consider two scenarios based on the primary mechanisms by which sGRBs are produced. The first one is the merger of two NSs, and according to the simulations, we use an amplified magnetic field of $10^{16}$ G. In contrast, in the second scenario, we consider $B=10^{12}$ G, which is the typical magnetic field intensity associated with the isolated NS during the BH-NS merger.
As we mentioned before, MeV-neutrinos are produced mainly by thermal processes. However, only $e^{\pm}$ capture on nuclei is the responsible one for the electronic neutrino production. So we have assumed a created neutrino proportion of $(\nu_e:\nu_\mu:\nu_\tau=4:3:3)$, i.e., for every ten created neutrinos, there will be four electronic, three muonic, and three tauonic neutrinos, respectively.\\

\begin{figure*}[htbp!] 
\centering
	\subfloat
					{  					\includegraphics[width=0.45\textwidth]{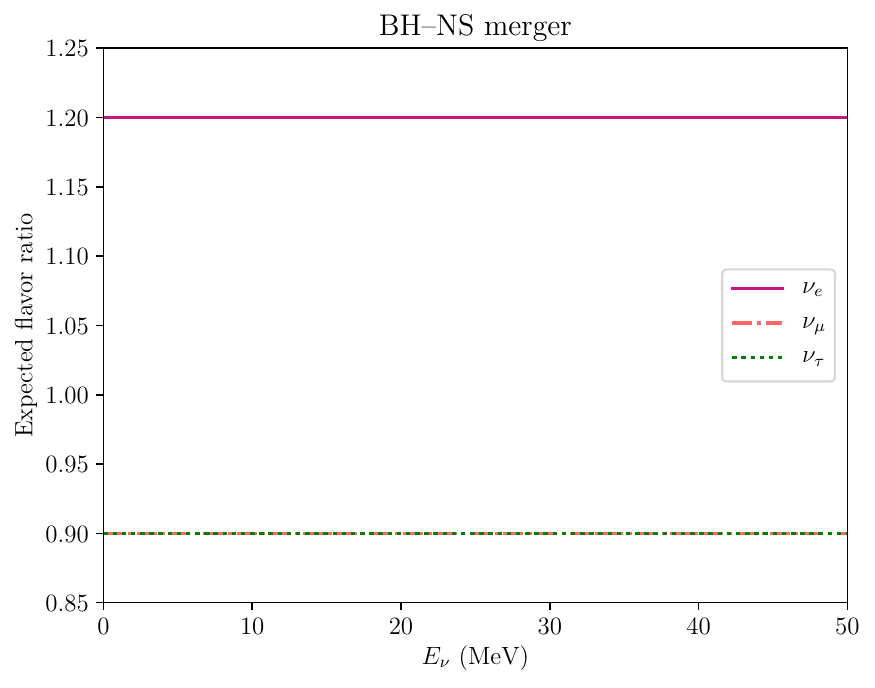}
 				    \label{plot:P_BH}
          	        }
					\qquad
	\subfloat
					{
  					\includegraphics[width=0.45\textwidth]{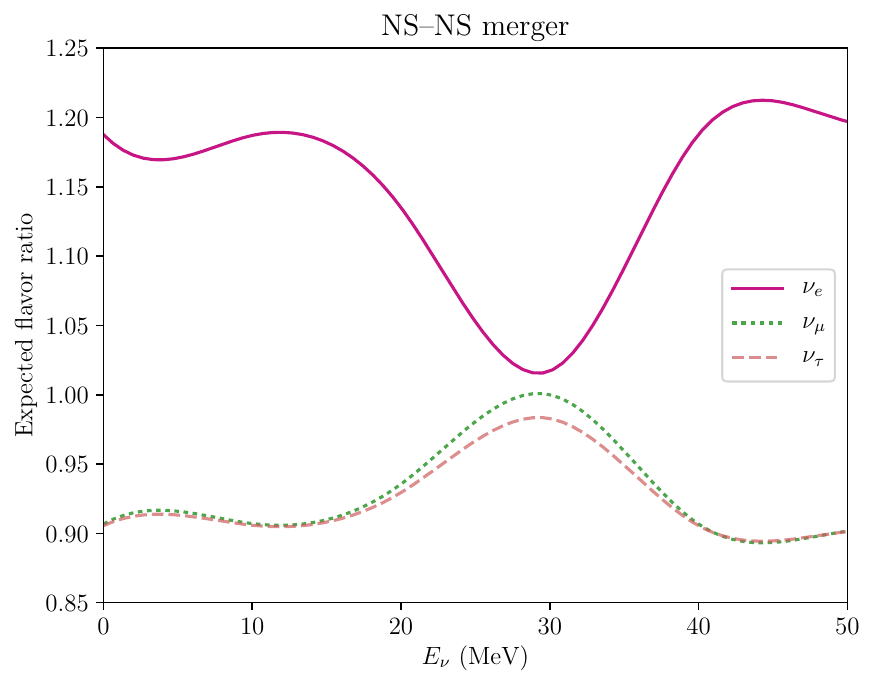}%
 				    \label{plot:P_BNS}
          	        }    
          	        \qquad  	       

\caption{Neutrino flavor ratio  expected on Earth as function of energy from a sGRB considering in the first case, a magnetic field of $B=10^{12}$ G (left panel) and in the second case a magnetic field of $B=10^{16}$ G (right panel).}
\label{probas1}
\end{figure*}

These neutrinos will cross a fireball medium before being released, and therefore the effective potential must be used to perform the propagation neutrino properties in both regimes.  It is important to notice that since neutrinos are already polarized into incoherent mass eigenstates after leaving the high-density source, then the neutrino oscillation phenomenon in a vacuum is suppressed \citep{lun01,rom15,smi05,kne08}.  For that reason, this effect will no longer be considered in this work. In this manner, the neutrino expected ratio on Earth will only be determined by the outgoing flavor ratio after neutrino oscillations occur within the source. Thus, considering both magnetic field scenarios, we first compute the transition probabilities from Equation  (\ref{ozzie}) using in the first case, the effective potential in the weak limit with $B=10^{12}$ G and in the second case, the potential in the strong limit with $B=10^{16}$ G. In both scenarios, we consider a medium with $r=10^7$ cm, $T=1$ MeV, $\mu=1$ keV and $\varphi=0^\circ$. These results allow us to obtain, in each case, the neutrino flavor ratio on Earth as a function of neutrino energies within MeV--range, which we show in Figure \ref{probas1}.\\

\begin{figure*}[h!] 
\centering
	\subfloat (a)
					{
  				\includegraphics[width=0.45\textwidth]{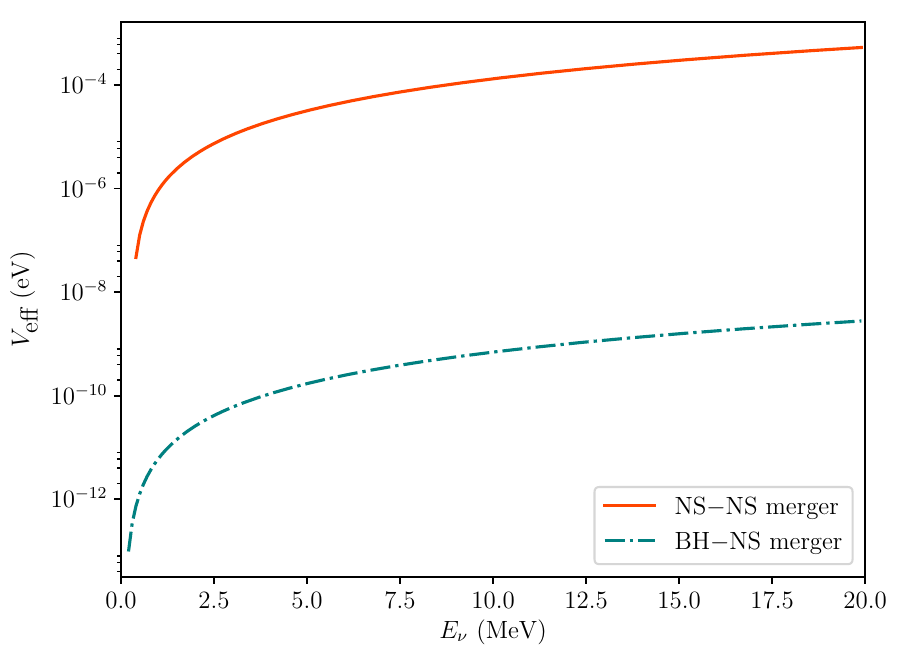}%
 				    \label{plot:V_Enu}
          	        }
	\subfloat (b) 
					{
  				\includegraphics[width=0.45\textwidth]{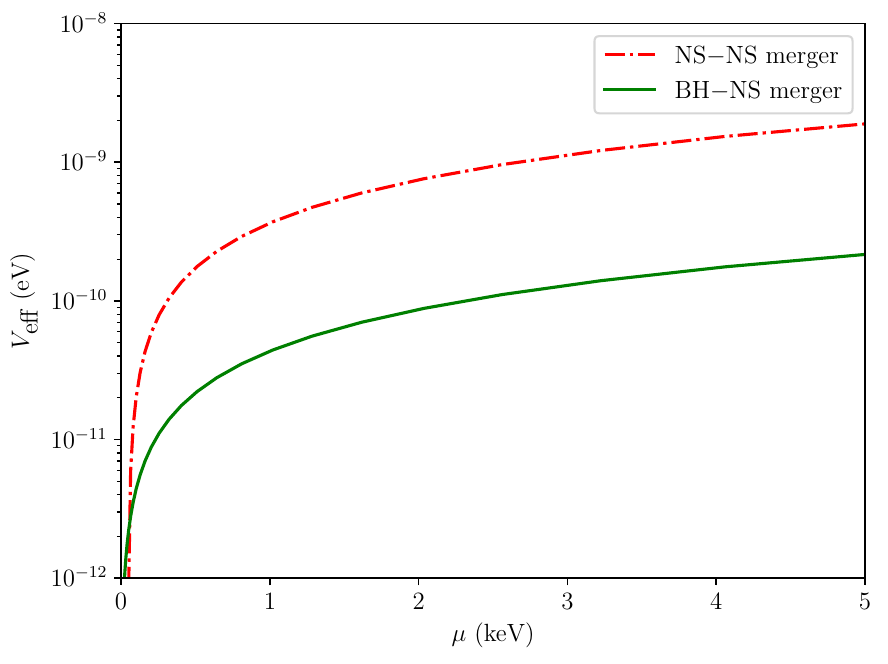}%
 				    \label{plot:V_mu}
          	        }    
          	        \qquad  	       
	\subfloat (c)
					{  					\includegraphics[width=0.45\textwidth]{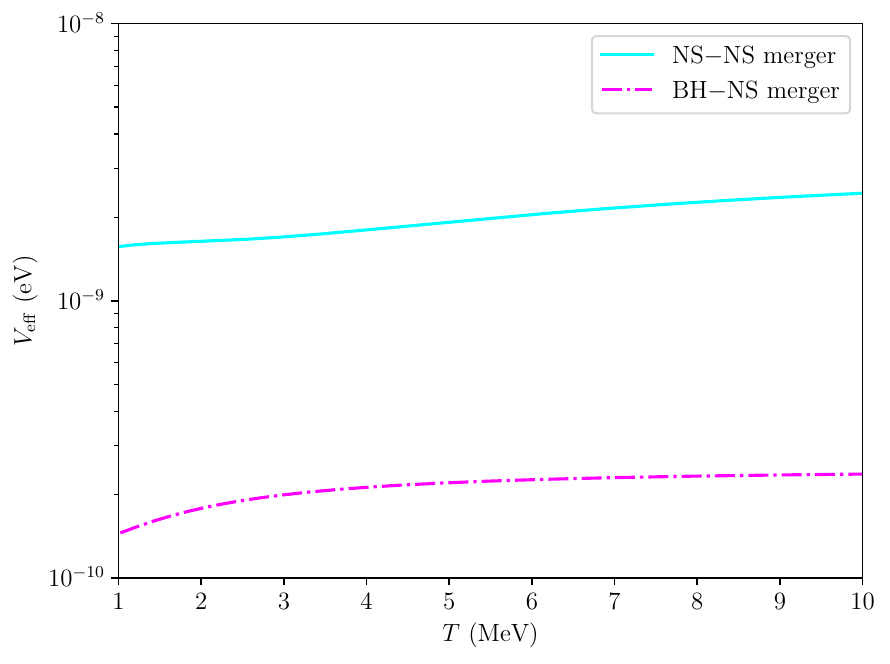}%
 				    \label{plot:V_T}
          	        }
	\subfloat (d)
					{
  				\includegraphics[width=0.45\textwidth]{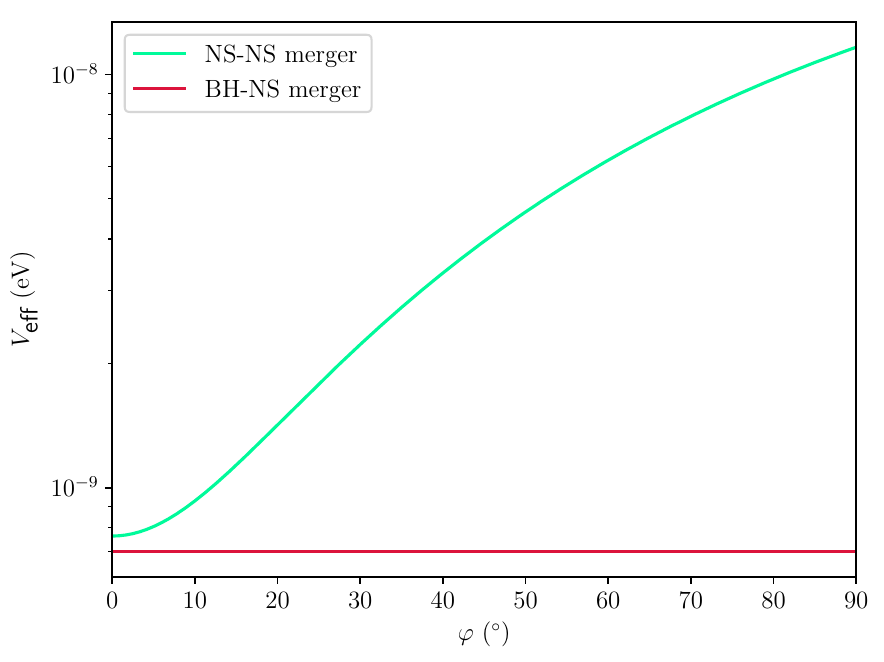}%
 				    \label{plot:V_phi}
          	        }   
\caption{Neutrino effective potential as a function of neutrino energy (a), chemical potential (b),  temperature (c) and neutrino propagation angle (d) , in both progenitor scenarios for multi-MeV neutrinos.  In subplot (d), we represent  the contribution to the potential using a magnetic field value of $B=10^{12}$ and   $B=10^{16}$ G. The magnetic field amplification present during a BNS merger evinces a substantial influence over the neutrino effective potential behavior.  }
\label{plot:Veff}
\end{figure*}

We see from Figure  \ref{probas1} that the same strong dependence of the magnetic field was found within the BNS merger scenario. On the one hand, the neutrino flavor ratio is fluctuating for several neutrino energies, while regarding a magnetic field of $B=10^{12}$ G, the ratio remains constant. This result suggests that if we detect MeV-neutrinos coming from the same source but with different flavor proportions at different energies, we can assure that the merger took place within an amplified magnetic field environment resulting from an NS-NS collision, but, on the other hand, detecting neutrinos with the same ratio within the whole treated energy interval prove that the magnetic field did not have any influence during the coalescence and, therefore, the central engine turns out to be a black hole and a neutron star. This behavior can be attributed to the direct influence that the magnetic field has over the effective neutrino potential. Thus, to notice in more detail those differences, we compute the potential in both regimes from Equation (\ref{eq:veffs}) and (\ref{eq:veffw}) shifting the involved physical parameters; these results are shown in Figure \ref{plot:Veff}. In each case, we have considered typical values lying within the range of $1\leq E_\nu\leq20$ MeV, $1\leq T\leq 10$ MeV and $1\leq\mu\leq1000$ eV. In this manner, we show in panel \ref{plot:Veff}a the contribution $V_{\rm eff}$ as a function of neutrino energy, while the chemical potential contribution is shown in panel \ref{plot:Veff}b. Furthermore, we exhibit the thermal contribution of the potential for i) a parallel propagation angle \ref{plot:Veff}c, and for ii) several propagation angles \ref{plot:Veff}d.  

We can notice that the potential increases to several orders of magnitude in those panels because of the magnetic field and propagation angle influence. The strong dependence on these variables suggests that these effects must be taken into account to understand the topology's internal magnetic field.

\begin{figure*}[htbp!] 
\centering
	\subfloat
					{
  					\includegraphics[width=0.48\textwidth]{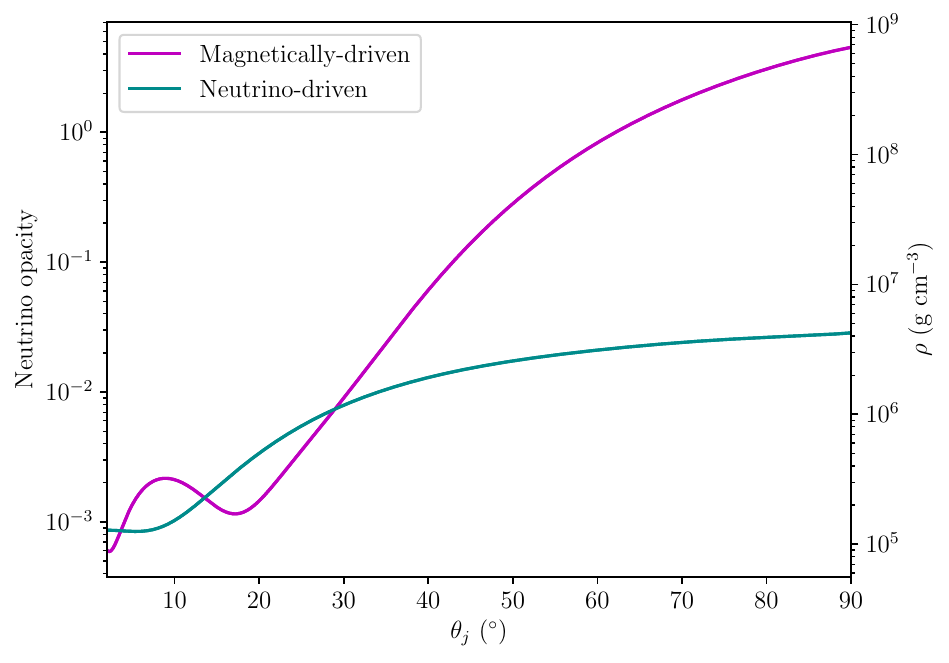}
 				    \label{plot:tau+rho}
          	        }
	\subfloat
					{
  					\includegraphics[width=0.45\textwidth]{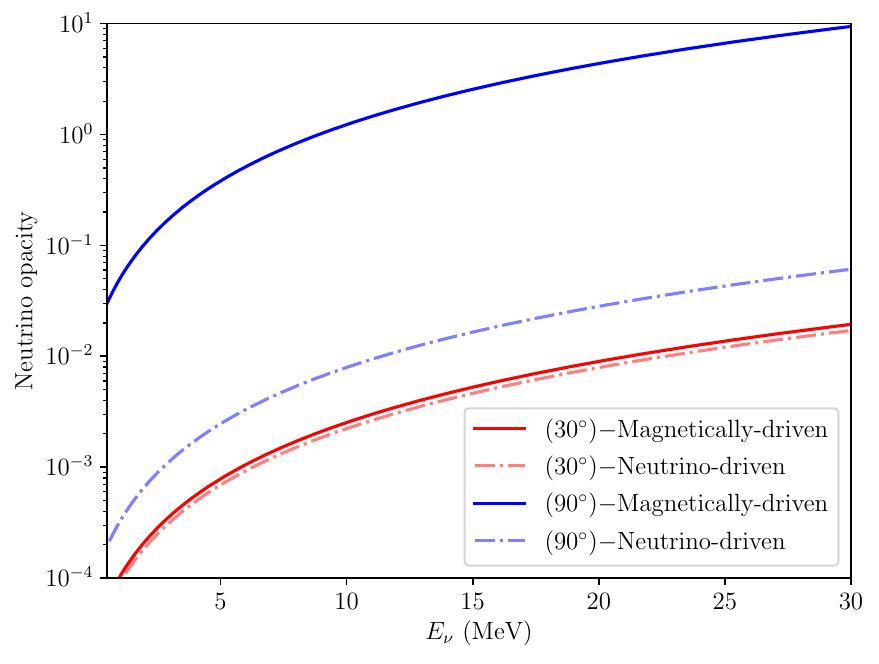}%
 				    \label{plot:tau_Enu}
          	        }              	     
\caption{ Left panel: Density profiles and neutrino opacity to the neutrino-driven and magnetically-driven winds  that surround the progenitor in both scenarios according with the analysis made by \cite{mur17} for a radius of $10^7$ cm, $E_\nu=20$ MeV, $Y_e=0.5$ and considering both neutrino processes  during propagation: neutrino capture on neutron (solid line) and antineutrino capture on protons (dashed line).\\
Right panel: Neutrino opacity in neutrino-driven  (dashed lines) and magnetically-driven winds (solid lines) for multiple half-opening angles and considering only neutrino capture processes. }
\label{plot:neutrino_opacity}
\end{figure*}

\subsection{Neutrino opacity: {on/off--axis} mechanism}

When neutrinos are released, they need to travel through the baryon-loaded winds produced during the merger \citep{ree98,sar00,zha02,dai04,rosII}. According to recent hydrodynamical (HD) and magneto-hydrodynamical (MHD) simulations, we realize that the medium density within these winds varies according to the type of progenitor involved \citep{des08,perego2014neutrino,siegel2014magnetically}.  Since the neutrino opacity  can be expressed as $\tau_{\nu(\bar{\nu})}=r/\lambda_\nu=\sigma_{\nu(\bar{\nu})} r n$, with $r$ the fireball radius and $n_e$ the electron  density, which is displayed in term of baryonic density as $n_e=Y_e \rho_{\rm w}(r,\theta_j)/m_p$, with $m_p$ the proton mass and $Y_e$ the average electron fraction. Then, the opacity turns out to be

\be \label{eq:tau_chida}  \tau_{\nu(\bar{\nu})}=\dfrac{\sigma_{\nu(\bar{\nu})} r Y_e \rho_{\rm w}(r,\theta_j)}{m_p^{-1}}\,, \ee\\
with $\sigma_{\nu(\bar{\nu})}$  obtained theoretically for MeV--neutrinos (antineutrinos) from \citep{tub75}:
 \begin{equation}\label{eq:sigma}
 \sigma_{\nu(\bar{\nu})}=\dfrac{G_F^2}{\pi} (3\alpha^2+1)E_\nu^2 g(E_\nu)\,,
 \end{equation}
with 
\be g(E_\nu)=\lp 1\pm \dfrac{Q}{E_\nu} \rp \lc 1\pm 2\dfrac{Q}{E_\nu} + \dfrac{Q^2-(\pm m_e^2)}{E_\nu^2} \rc^{1/2}\,,
\ee

here the signus $\pm$ corresponds  to the neutrino (+) and antineutrino (-) treatment, $Q=1.3$ MeV the proton-neutron mass difference, $\alpha=-1.26$ the nuclear axial coupling coefficient. Then, the neutrino cross--section  lies in the range between $10^{-43}$ and $10^{-40}$ cm$^2$, for the multi-MeV energy range.\\


To estimate the number of neutrinos released in an extreme event, such as a GRB, mostly, we need to compute the baryon density profiles $\rho(\theta_j)$ of the wind-like circumstellar media. Recent hydrodynamical (HD) and magneto-hydrodynamical (MHD) simulations have been performed to achieve this purpose \citep{perego2014neutrino,siegel2014magnetically}.  \cite{mur17} compiled this result in both regimes, finding the density profiles shown in Figure \ref{plot:neutrino_opacity}, where we can observe a considerable increment in density for angles close to the equatorial plane due to the magnetic field amplification. Once we know the density profiles and neutrino cross-section, from Equation (\ref{eq:tau_chida}), we have plotted the neutrino opacity around the progenitor originated by circumburst winds as a function of the half-opening angle $\theta_j$ and neutrino energy, respectively. In these plots, the most remarkable result is the strong dependence on the propagation during a BNS merger. For instance, we observe that the medium is opaque for 20-MeV neutrinos propagating at $\theta_j>62.1^\circ$, as well as for 30-MeV neutrinos propagating at $\theta_j=54.1^\circ$, and indeed,  considering extreme energies like $E_\nu=100$ MeV, we find that neutrinos propagating at $\theta_j>38.2^\circ$ will not be released at all. By comparison, during a BH--NS coalescence, neutrinos are isotropically released for the whole MeV-energy range. The above directly implies that where a magnetic field amplification effect occurs, neutrinos are confined in a preferential direction along the jet propagation path.  Therefore, we can be able to distinguish between central engines if, for example, we identify the electromagnetic counterpart of an off-axis GRB and instead; we do not detect neutrinos from the same source, being this a plausible method to discriminate among both kinds of progenitors (BH-NS and NS-NS).

\section{Detection}\label{sec:detection}

\subsection{Detectors}

\begin{figure*}[htbp!] 
\centering
	\subfloat
					{
  					\includegraphics[width=0.45\textwidth]{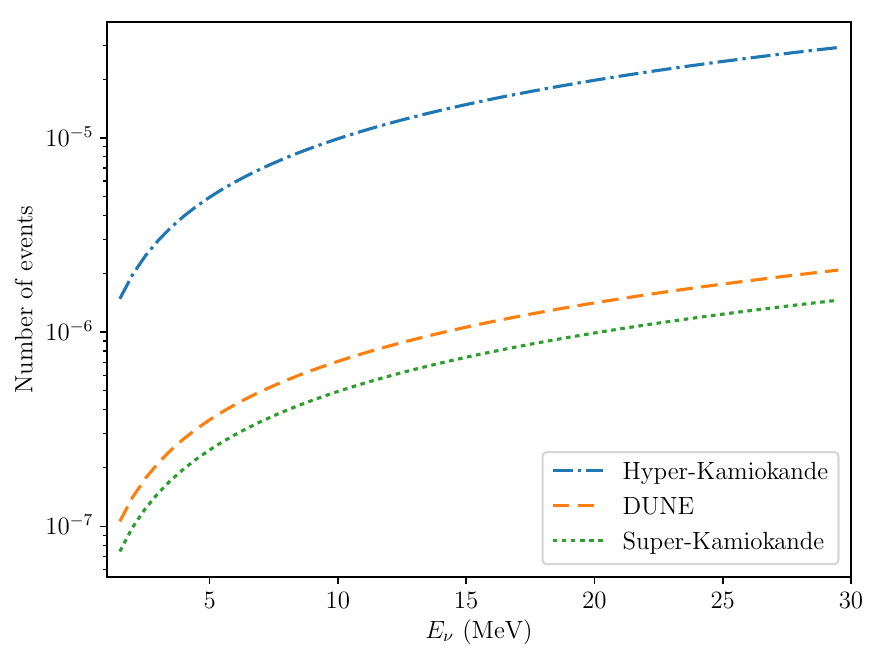}
 				    \label{plot:Nev_E}
          	        }
	\subfloat
					{
  				\includegraphics[width=0.45\textwidth]{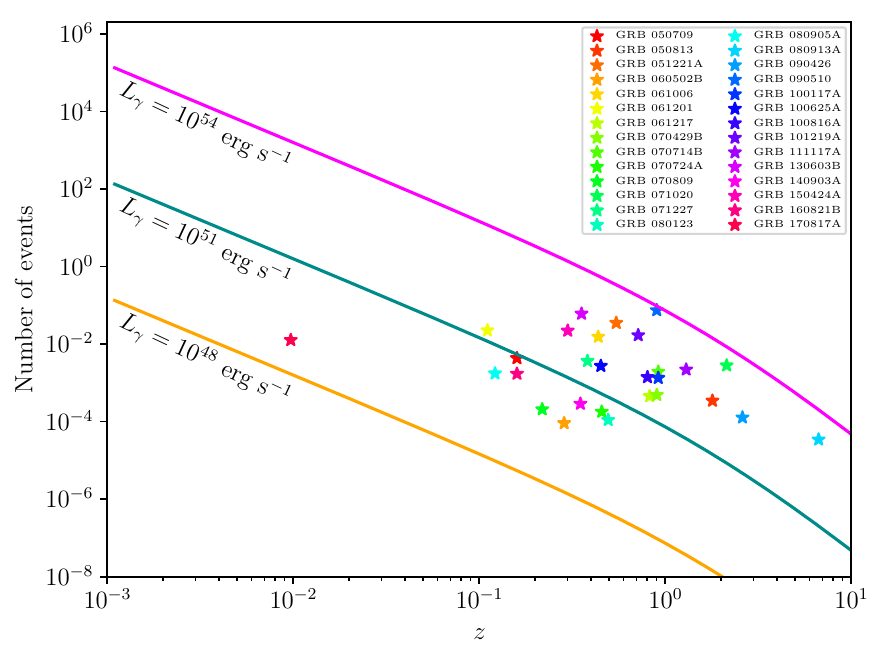}%
 				    \label{plot:Nev_sGRB}
          	        }

\caption{\\
Left panel: Number of neutrino expected events on several ground-based neutrino telescopes as a function of typical thermal neutrino energies $E_\nu$ using as a case study the GW170817/GRB170817A event. In this plot, we have considered the values of $L_{\nu}=5\times10^{48}$ erg s$^{-1}$, $d_z=40$ Mpc and $T_{\text{burst}}=2$ s.\\
Right panel: Number of neutrino expected events from sGRBs in Hyper-Kamiokande detector as a redshift function considering an average neutrino energy of  $E_\nu=20$ MeV. Additionally, we compare the expected neutrino events for several sGRB sources with known redshift and isotropic-equivalent luminosities.}
\label{Neventos}
\end{figure*}

\subsubsection{Super-Kamiokande}
Super-Kamiokande (SK) is an underground neutrino observatory built 1000 m below the surface in a Japanese mine. At present, SK is a giant water  \v{C}erenkov detector with a base of 39 m in diameter and a height of 42 m, having the capacity to contain 50 kton of ultra-pure water. SK has an array of more than 13,000 photomultipliers (PMTs) placed in two sections;  11,129  within the inner region and 1,885 PMTs in the outer region. The SK outreach performs studies in the solar, accelerator, and atmospheric neutrinos and proton decay in the MeV energy range \citep{fuk03}.
\subsubsection{Hyper-Kamiokande}

Hyper-Kamiokande (HK) will be a third-generation \v{C}erenkov detector (replacing the current SK detector) also located in a Japanese mine near to its predecessor SK. It is expected to start operating somehow in the middle of this decade.  The original design \citep{2011arXiv1109.3262A,hk14} shows a design of two almost-cylindrical tanks containing (0.56) million metric tons of ultra-pure water. This detector will be equipped with 99,000 PMTs uniformly placed within the tanks. Complementing the tasks of SK, HK will also perform detection for neutrinos produced in both terrestrial and extra-terrestrial sources, as well as studies in  Particle Physics, such as CP violation in the leptonic sector, proton decay, and neutrino oscillation phenomena \citep{hk18}.

\subsubsection{Deep Underground Neutrino Experiment}

The DUNE (\textit{Deep Underground Neutrino Experiment}) experiment will consist of two neutrino experiments, the first one placed near to Fermi National Laboratory Acceleration Facility in Illinois, USA (short-baseline program). In contrast, the second one will be built within the SURF facilities in South Dakota (long-baseline program). Altogether DUNE will have a fiducial mass of 40 kton of liquid argon, within four cryostats adapted with Liquid Argon Time
Projection Chambers (LArTPCs) and analogously to its counterpart (HK) it is expected to start operating at the end of this decade. 
Their primary focus will be the studies of accelerator neutrinos and the measure of its mixing parameters. However, it will also perform studies to detect astrophysical neutrinos and the search for proton decay \citep{acc16}.

\subsection{Number of neutrino expected events}

It is possible to estimate the numbers of events to be expected on Earth as\\
%
%
\be
N_{\rm ev}=V N_A\,  \rho_N  \int_{t'} \int_{E'} \sigma^{\bar{\nu}_ep}_{cc} \frac{dN}{dE}\,dE dt\,,
\ee
\noindent where $V$ is the effective volumen of water, $N_A=6.022\times 10^{23}$ g$^{-1}$ is the Avogadro's number, $\rho_N=(M_{\rm fiducial}/ V)=2/18\, {\rm g\, cm^{-3}}$ is the nucleons density in water or $ \sigma^{\bar{\nu}_ep}_{cc}\simeq 9\times 10^{-44}\,E^2_{\bar{\nu}_e}/{\rm MeV}^2$  is the cross-section  \citep{1989neas.book.....B},   $dt$ is the neutrino emission time and $dN/dE$ is the neutrino spectrum. Taking into account the relation between the  neutrino luminosity $L_{\bar{\nu}_e}$ and flux $F_{\bar{\nu}_e}$, $L_{\bar{\nu}_e}=4\pi d^2_z  F_{\bar{\nu}_e}\braket{E_{\bar{\nu}_e}}=4\pi d^2_z   E^2_{\bar{\nu}_e} dN/dE_{\bar{\nu}_e}$, then the number of events is 
\bary\label{num_Neu}
N_{\text{ev}}&\simeq& \frac{N_A\,  \rho_N  \sigma^{\bar{\nu}_ep}_{cc}}{4\pi d^2_z \braket{E_{\bar{\nu}_e}}}V_{\rm det} \,\braket{E_{\bar{\nu}_e}}\,,
\eary
where $d_z$ is the distance from neutrino production to Earth, $\braket{E_{\bar{\nu}_e}}$ is the average energy of electron antineutrino and $E_{T,\bar{\nu}_e}=\int L_{\bar{\nu}_e} dt$  is the total energy emitted \citep{2004mnpa.book.....M, 2014MNRAS.442..239F}. \\

The number of neutrino expected events can be obtained through Equation (\ref{num_Neu}). As a case of study we compute the number of plausible neutrino events coming from GRB170817, for this purpose we regard the reported initial parameters for this event with $L_{\nu}=5\times10^{48}$ erg s$^{-1}$, $d_z=40$ Mpc and $T_{\text{burst}}=2$ s \citep{abott17}. We represent in the left panel of Figure \ref{Neventos} the expected events in {HK, DUNE}, and {SK}  detectors as a function of neutrino energy.   We found that even though this GRB was in a nearby location (40 Mpc), it had an atypical low isotropic luminosity that diminishes the initial neutrino flux. Multiple observations performed for several collaborations in different energy scales \citep{alb17} have verified that no neutrino signal has been detected on Earth from this source.\\

We realize that HK detection capacity is more than an order of magnitude greater than {DUNE} experiment with its predecessor SK.   Being HK, the best candidate so far to perform extragalactic neutrino detections, we present in the right panel of Figure \ref{Neventos} the number of neutrinos expected events in HK detector collecting data information for the most significant sGRBs with a measured redshift thus far \citep{berger2014short,dav14,jin18}. In our calculations, we have considered an Einstein--de Sitter (EdS) Universe with  $h=0.673$,   $\Omega_m=0.315$, $\Omega_\Lambda=0.685$ parameters, being  $\Omega_m$ and $\Omega_\Lambda$ the pressureless matter density and the dark energy density of the $\rm{\Lambda CDM}$ Universe \citep{patrignani2016review}, respectively. We also notice that most of the sGRBs have their origin in a host galaxy with $z>1$ and have a typical equivalent isotropic luminosity between $10^{51}-10^{53}$ erg s$^{-1}$, being GRB 090524 the brightest one, almost reaching the value of $10^{54}$ erg s $^{-1}$. In the same plot, we can see the atypical GRB associated with the GW detection in 2017, coming to be the least luminous and the closest GRB in this sample. Additionally, we recognize the most plausible sensitive detection region in HK for multi--MeV neutrinos produced in sGRBs. We can see that, for instance, a close event like GRB 170817A  but with a luminosity tantamount to a typical GRB could lead to greater neutrino flux. In this case, the HK experiment could detect them.

\section{Conclusion}\label{sec:conclusion}

We can conclude that multi-MeV neutrinos play an important role in understanding transient events where radiation cannot break free during the early phases. In that context, we studied the effect on the neutrino effective potential product of a magnetic field amplification during the merger of both kinds of progenitors that give rise to sGRBs. We found that even when the potential is dependant on multiple physical parameters, only the magnetic field intensity, and the neutrino propagation angle represent the greatest contribution to the potential, coming to be up to four orders of magnitude bigger in the NS-NS scenario. It becomes crucial because, through this outcome, we can infer the intrinsic topology composition of the involved magnetic field during the merger. On the other hand, to figure out how the properties of the neutrinos are altered, we have computed the neutrino oscillations and resonance lengths in both magnetic field regimes for these multi-MeV neutrinos.  We found that they will oscillate resonantly before leaving the fireball with a predominant $\nu_e$ survival rate affecting the final expected neutrino ratio on Earth. After computing the neutrino probabilities, we realized that we could discriminate against the progenitors based on the flavor ratio expected through this effect. For instance, taking two arbitrary energy values during a BNS merger, we expect a flavor ratio of $(\nu_e:\nu_\mu:\nu_\tau=  1.1871	:0.9071:	0.9059)$ for $E_\nu=10 $ MeV and  $(\nu_e:\nu_\mu:\nu_\tau=  1.0171	:1.000:	0.9829)$ for $E_\nu=30$ MeV,  while for a BH--NS merger, we will expect the same ratio of $(\nu_e:\nu_\mu:\nu_\tau=  1.2:0.9:0.9)$ for both chosen energies

Concerning the released neutrinos, we identified two different behaviors in both configurations: i) in an NS-NS merger; neutrinos are collimated along the rotation axis, mainly because of the gradual angular decrease of the opacity within the surrounding medium during the magnetic field amplification, and ii) during a BH--NS merger; the medium remains transparent towards neutrinos, regardless the angle by which they leave the source, meaning that neutrinos are isotropically released. This result provides a unique, trustworthy method to discriminate against the participating progenitors during the merger through this \textit{on-axis/off-axis} method.   Since if somehow we can detect an off-axis GRB with a line of sight greater than the critical angle (for instance, through their afterglow), and jointly we  detect neutrinos within temporal and spatial dependency associated with this source, then we could determine the GRB progenitors analyzing the dynamics associated when the magnetic field amplification took place. 

Our estimates also predict that neutrinos from an energetic sGRB with luminosity $(L\gtrsim 10^{52}$ erg s$^{-1})$ located within a nearby vicinity such as GRB 170817A ($d_z=40$ Mpc) could be detected in the HK detector. Nevertheless, no multi-MeV neutrino has been found so far from sources with the characteristics presented above, which agrees with our results.

\newpage
\appendix
\begin{widetext}

\subsubsection{Strong $\vec{B}$ limit}

In the strong magnetic field  approximation ($ B/B_c \gg 1$), the Lorentz scalars are
{\small
\bary\nonumber
a_{\rm \perp, s}=-\frac{\sqrt2G_F}{m_W^2}\biggl[ \biggl( (E_{\nu_e}+ k_3) (N^0_e - \bar{N}^0_e) \biggr) +\frac{eB}{\pi^2}\int^\infty_0 dp_3 \,\frac{m_e^2}{E_0}\, (f_{e,0}+\bar{f}_{e,0})\biggr],
\label{astr}
\eary
}
{\small
\bary\label{b_s}\nonumber
b_{\rm s}=&&\sqrt2 G_F\biggl[ \biggl( 1+\frac32\frac{m_e^2}{M_W^2}+ \frac{eB}{M_W^2} +\frac{E_{\nu_e}k_3}{M_W^2}+\frac{E^2_{\nu_e}}{M_W^2}\biggr)(N^0_e - \bar{N}^0_e)-\frac{eB}{2\pi^2M_W^2}\int^\infty_0\,dp_3\biggl\{2\,k_3E_{e,0}+2E_{\nu_e}\biggl( E_{e,0} - \frac{m_e^2}{2E_{e,0}}\biggr) \biggr\} \cr
&&\hspace{13.9cm} \times (f_{e,0}+\bar{f}_{e,0})       \biggr]\,,
\label{Lescb}
\eary
}
and
{\small
\bary\label{c_s}
c_{\rm s}=&&\sqrt2 G_F\biggl[ \biggl( 1+\frac12\frac{m_e^2}{M_W^2}+ \frac{eB}{M_W^2} -\frac{E_{\nu_e}k_3}{M_W^2}-\frac{k^2_3}{M_W^2}\biggr)(N^0_e - \bar{N}^0_e)-\frac{eB}{2\pi^2M_W^2}\int^\infty_0\,dp_3\biggl\{ 2E_{\nu_e}\biggl( E_{e,0} - \frac{m_e^2}{2E_{e,0}}\biggr) \biggr\}\cr
&&\hspace{9.9cm} +2k_3\biggl( E_{e,0} - \frac{3m_e^2}{2E_{e,0}}\biggr) \biggr\}  (f_{e,0}+\bar{f}_{e,0})       \biggr]\,,
\label{Lescc}
\eary
}
where the number density of electrons can be written as {\small $n_e^0(\mu, T, B)=\frac{eB}{2\pi^2}\int^\infty_0 dp_3 f_{e,0}$} with {\small $f_{e,0}=f(E_{e,0})=\frac{1}{e^{\beta(E_{e,0}-\mu)} +1}$} and  {\small $E^2_{e,0}=(p^2_3+m_e^2)$}.  The quantity $E_0$ corresponds to the  electron energy in the lowest Landau level.\\
Assuming that  the chemical potentials ($\mu$) of the electrons and positrons are much smaller than their energies ($\mu\leq$E$_{e,0}$), the fermion distribution function can be written as a sum  given by {\small $f_{e,0} \approx\sum^{\infty}_{l=0}(-1)^l e^{-\beta(E_{e,0}-\mu)(l+1)}$}.\\
%
%
Replacing Equations (\ref{c_s}) in (\ref{poteff}) and solving these integral-terms,  the neutrino effective potential in the strong magnetic field regime becomes
\begin{eqnarray}\label{Veffs}
V_{eff,s}=\frac{\sqrt2\,G_F\,m_e^3 B}{\pi^2\,B_c}\biggr[\sum^{\infty}_{l=0}(-1)^l\sinh\alpha_l   \left[F_s-G_s\cos\varphi \right] -4\frac{m^2_e}{m^2_W}\,\frac{E_\nu}{m_e}\sum^\infty_{l=0}(-1)^l\cosh\alpha_l  \left[J_s-H_s\cos\varphi \right]  \biggr]\,,
 \end{eqnarray}
where $m_e$ is the electron mass, $\alpha_l=(l+1)\mu/T$  with $\mu$ and $T$ the chemical potential and temperature, respectively, $B_c=4.141\times 10^{14}\, {\rm G}$ is the critical magnetic field,  $E_\nu$ is the neutrino energy and the functions F$_s$, G$_s$, J$_s$, H$_s$ are
\bary
F_s&=&\left[1+ \frac{m_e^2}{m^2_W}\left(\frac32+2\frac{E^2_\nu}{m^2_e} +\frac{B}{B_c}\right)\right]K_1(\sigma_l)\,, \hspace{1cm}  G_s=\left[1+ \frac{m_e^2}{m^2_W}\left(\frac12-2\frac{E^2_\nu}{m^2_e} +\frac{B}{B_c}\right)\right]K_1(\sigma_l)\,,\cr
J_s&=&\frac34K_0(\sigma_l)+\frac{K_1(\sigma_l)}{\sigma_l} \,, \hspace{3.8cm}    H_s=\frac{K_1(\sigma_l)}{\sigma_l}\,,
\eary
with $\sigma_l=(l+1)m_e/T$.
\subsubsection{Weak $\vec{B}$ limit}
%
%
In the weak field approximation ($ B/B_c \ll 1$),   the potential in this regimen can be written as 

{\small
\bary\nonumber
a_{\rm \perp, w}=-\frac{\sqrt2G_F}{m_W^2}\biggl[ \biggl\{E_{\nu_e}(n_e-\bar{n}_e)+ k_3(n_e^0-\bar{n}_e^0)\biggr\}+\frac{eB}{2\pi^2}\int^\infty_0 dp_3  \int^\infty_0    (2-\delta_{n,0}) \biggl (\frac{m_e^2}{E_n}- \frac{2neB}{E_n}\biggr)(f_{e,n}+\bar{f}_{e,n})\,dn \biggr],
\label{conaw}
\eary
}
{\small
\bary\nonumber
b_{\rm w}&=&\sqrt2 G_F \biggl[\biggl(1+\frac{E_{\nu_e}^2}{m_W^2}\biggr)(n_e-\bar{n}_e)+ \frac{ E_{\nu_e}k_3}{m_W^2}(n_e^0-\bar{n}_e^0)-\frac{eB}{\pi^2m_W^2} \int^\infty_0 dp_3 \int^\infty_0  (2-\delta_{n,0})E_{\nu_e}\biggl\{E_n\delta_{n,0} +\biggl(E_n -\frac{m_e^2}{2E_n}\biggr)\biggr\}\cr
&&\hspace{12.9cm} \times(f_{e,n}+\bar{f}_{e,n})\,dn \biggr]\,, \nonumber
\label{conbw}
\eary
}
and
{\small
\bary
c_{\rm w}=\sqrt2 G_F\biggl[\biggl(1-\frac{k_3^2}{m_W^2}\biggr)(n_e^0-\bar{n}_e^0) -\frac{E^2_{\nu_e}}{m_W^2}(n_e-\bar{n}_e) -\frac{eB}{\pi^2m_W^2} \int^\infty_0 dp_3\int^\infty_0  (2-\delta_{n,0}) E_{\nu_e}\biggl\{\biggl(E_n-\frac{m_e^2}{E_n}\biggr)\delta_{n,0}\nonumber \\
+\biggl(E_n-\frac32\frac{m_e^2}{E_n}-\frac{2neB}{E_n}\biggr)\biggr\}(f_{e,n}+\bar{f}_{e,n})\, dn \biggr].
\label{concw}
\eary
}

where the electron number density is {\small $n_e(\mu, T, B)=\frac{eB}{2\pi^2}    \int_0^\infty  dp_3  \int^\infty_0    \frac{(2 - \delta_{n,0})\, dn}{e^{\beta(E_{e,n}-\mu)} +1}$} and electron distribution function is {\small $f_{e,n}=f(E_{e,n},\mu)=\frac{1}{e^{\beta(E_{e,n}-\mu)} +1}$}, with $\bar{f}_{e,n}(\mu, T)=f_{e,n}(-\mu, T)$ and {\small$E_{e,n}={p_3^2+m_e^2+2neB}^{1/2}$}.\\
Replacing Equations (\ref{concw}) in (\ref{poteff}) and solving these integral-terms,  the neutrino effective potential in the strong magnetic field regime becomes
\begin{eqnarray}\label{Veffw}
V_{eff,is(w)}=\frac{\sqrt2\,G_F\,m_e^3 B}{\pi^2\,B_c}\biggr[\sum^{\infty}_{l=0}(-1)^l\sinh\alpha_l  \left[F_w-G_w\cos\varphi \right]-4\frac{m^2_e}{m^2_W}\,\frac{E_\nu}{m_e}\sum^\infty_{l=0}(-1)^l\cosh\alpha_l \left[J_w-H_w\cos\varphi \right] 
\end{eqnarray}
where the functions F$_w$, G$_w$, J$_w$, H$_w$ are 
\bary
F_w&=&\biggl(2+2\frac{E^2_\nu}{m^2_W}\biggr) \biggl(\frac{K_0(\sigma_l)}{\sigma_l}+2\frac{K_1(\sigma_l)}{\sigma_l^2} \biggr) \frac{B_c}{B}-K_1(\sigma_l)\,,  \hspace{1cm} G_w=K_1(\sigma_l)-\frac{2B_c}{B}\frac{E^2_\nu}{m^2_W}    \biggl(\frac{K_0(\sigma_l)}{\sigma_l}+2\frac{K_1(\sigma_l)}{\sigma_l^2}\biggr)\,, \nonumber\\
J_w&=&\biggl(\frac12+\frac{3B_c}{B\,\sigma_l^2}\biggr)K_0(\sigma_l)+\frac{B_c}{B}  \biggl(1+\frac{6}{\sigma_l^2}\biggr)     \frac{K_1(\sigma_l)}{\sigma_l}\,,  \hspace{1.6cm}   H_w= \biggl(\frac12+\frac{B_c}{B\,\sigma_l^2} \biggr)K_0(\sigma_l)+\frac{B}{B_c} \biggl(\frac{2}{\sigma_l^2}-\frac12\biggr)\frac{K_1(\sigma_l)}{\sigma_l}\,.\nonumber\\
\eary

\acknowledgements
We thank J. Beacom, C. G. Bernal, and D. Page for useful discussions.  This work was supported by  PAPIIT-UNAM IA102019 and IN113320.

\end{widetext}

\bibliographystyle{unsrt85}
\bibliography{Bib/Bib_osc}

\begin{thebibliography}{100}

\bibitem{SN1987A}
K.~Hirata, T.~Kajita, M.~Koshiba, et~al.
\newblock Observation of a neutrino burst from the supernova sn1987a.
\newblock {\em Phys. Rev. Lett.}, 58:1490--1493, Apr 1987.

\bibitem{GW150914}
B.~P. {Abbott}, R.~{Abbott}, T.~D. {Abbott}, et~al.
\newblock {Binary Black Hole Mergers in the First Advanced LIGO Observing Run}.
\newblock {\em Physical Review X}, 6(4):041015, October 2016.

\bibitem{abott17}
Benjamin~P Abbott, R~Abbott, TD~Abbott, et~al.
\newblock Gw170817: observation of gravitational waves from a binary neutron
  star inspiral.
\newblock {\em Physical Review Letters}, 119(16):161101, 2017.

\bibitem{alb17}
A.~Albert, M.~Andr{\'{e}}, M.~Anghinolfi, et~al.
\newblock Search for high-energy neutrinos from binary neutron star merger
  {GW}170817 with {ANTARES}, {IceCube}, and the pierre auger observatory.
\newblock {\em The Astrophysical Journal}, 850(2):L35, nov 2017.

\bibitem{att17}
J.-L. Atteia, V.~Heussaff, J.-P. Dezalay, et~al.
\newblock The maximum isotropic energy of gamma-ray bursts.
\newblock {\em The Astrophysical Journal}, 837(2):119, mar 2017.

\bibitem{kle73}
R.~W. {Klebesadel}, I.~B. {Strong}, and R.~A. {Olson}.
\newblock {Observations of Gamma-Ray Bursts of Cosmic Origin}.
\newblock {\em \apjl}, 182:L85, June 1973.

\bibitem{berger2014short}
Edo Berger.
\newblock Short-duration gamma-ray bursts.
\newblock {\em Annual Review of Astronomy and Astrophysics}, 52:43--105, 2014.

\bibitem{2004IJMPA..19.2385Z}
B.~{Zhang} and P.~{M{\'e}sz{\'a}ros}.
\newblock {Gamma-Ray Bursts: progress, problems and prospects}.
\newblock {\em International Journal of Modern Physics A}, 19:2385--2472, 2004.

\bibitem{2015PhR...561....1K}
P.~{Kumar} and B.~{Zhang}.
\newblock {The physics of gamma-ray bursts and relativistic jets}.
\newblock {\em \physrep}, 561:1--109, February 2015.

\bibitem{hjo03}
J.~{Hjorth} and et~al.
\newblock {A very energetic supernova associated with the {$\gamma$}-ray burst
  of 29 March 2003}.
\newblock {\em \nat}, 423:847--850, June 2003.

\bibitem{woo06}
S.~E. {Woosley} and J.~S. {Bloom}.
\newblock {The Supernova Gamma-Ray Burst Connection}.
\newblock {\em \araa}, 44:507--556, September 2006.

\bibitem{hjo12}
J.~{Hjorth} and J.~S. {Bloom}.
\newblock {\em {The Gamma-Ray Burst - Supernova Connection}}, pages 169--190.
\newblock November 2012.

\bibitem{eichler1989nucleosynthesis}
David Eichler, Mario Livio, Tsvi Piran, and David~N Schramm.
\newblock Nucleosynthesis, neutrino bursts and $\gamma$-rays from coalescing
  neutron stars.
\newblock {\em Nature}, 340(6229):126, 1989.

\bibitem{lee04}
W.~H. {Lee}, E.~{Ramirez-Ruiz}, and D.~{Page}.
\newblock {Opaque or Transparent? A Link between Neutrino Optical Depths and
  the Characteristic Duration of Short Gamma-Ray Bursts}.
\newblock {\em \apjl}, 608:L5--L8, June 2004.

\bibitem{lee05}
W.~H. {Lee}, E.~{Ramirez-Ruiz}, and J.~{Granot}.
\newblock {A Compact Binary Merger Model for the Short, Hard GRB 050509b}.
\newblock {\em \apjl}, 630:L165--L168, September 2005.

\bibitem{lee2007progenitors}
William~H Lee and Enrico Ramirez-Ruiz.
\newblock The progenitors of short gamma-ray bursts.
\newblock {\em New Journal of Physics}, 9(1):17, 2007.

\bibitem{nak07}
E.~{Nakar}.
\newblock {Short-hard gamma-ray bursts}.
\newblock {\em \physrep}, 442:166--236, April 2007.

\bibitem{zra13}
J.~{Zrake} and A.~I. {MacFadyen}.
\newblock {Magnetic Energy Production by Turbulence in Binary Neutron Star
  Mergers}.
\newblock {\em \apjl}, 769:L29, June 2013.

\bibitem{Price719}
D.~J. Price and S.~Rosswog.
\newblock Producing ultrastrong magnetic fields in neutron star mergers.
\newblock {\em Science}, 312(5774):719--722, 2006.

\bibitem{gia09}
B.~{Giacomazzo}, L.~{Rezzolla}, and L.~{Baiotti}.
\newblock {Can magnetic fields be detected during the inspiral of binary
  neutron stars?}
\newblock {\em \mnras}, 399:L164--L168, October 2009.

\bibitem{obe10}
M.~{Obergaulinger}, M.~A. {Aloy}, and E.~{M{\"u}ller}.
\newblock {Local simulations of the magnetized Kelvin-Helmholtz instability in
  neutron-star mergers}.
\newblock {\em \aap}, 515:A30, June 2010.

\bibitem{kiu14}
Kenta Kiuchi, Koutarou Kyutoku, Yuichiro Sekiguchi, Masaru Shibata, and
  Tomohide Wada.
\newblock High resolution numerical relativity simulations for the merger of
  binary magnetized neutron stars.
\newblock {\em Phys. Rev. D}, 90:041502, Aug 2014.

\bibitem{kiu15}
K.~{Kiuchi}, P.~{Cerd{\'a}-Dur{\'a}n}, K.~{Kyutoku}, Y.~{Sekiguchi}, and
  M.~{Shibata}.
\newblock {Efficient magnetic-field amplification due to the Kelvin-Helmholtz
  instability in binary neutron star mergers}.
\newblock {\em \prd}, 92(12):124034, December 2015.

\bibitem{2010Natur.467.1081D}
P.~B. {Demorest}, T.~{Pennucci}, S.~M. {Ransom}, M.~S.~E. {Roberts}, and
  J.~W.~T. {Hessels}.
\newblock {A two-solar-mass neutron star measured using Shapiro delay}.
\newblock {\em \nat}, 467:1081--1083, October 2010.

\bibitem{2013Sci...340..448A}
J.~{Antoniadis}, P.~C.~C. {Freire}, N.~{Wex}, et~al.
\newblock {A Massive Pulsar in a Compact Relativistic Binary}.
\newblock {\em Science}, 340:448, April 2013.

\bibitem{1992ApJ...392L...9D}
R.~C. {Duncan} and C.~{Thompson}.
\newblock {Formation of very strongly magnetized neutron stars - Implications
  for gamma-ray bursts}.
\newblock {\em \apjl}, 392:L9--L13, June 1992.

\bibitem{2008MNRAS.385.1455M}
B.~D. {Metzger}, E.~{Quataert}, and T.~A. {Thompson}.
\newblock {Short-duration gamma-ray bursts with extended emission from
  protomagnetar spin-down}.
\newblock {\em \mnras}, 385:1455--1460, April 2008.

\bibitem{1977MNRAS.179..433B}
R.~D. {Blandford} and R.~L. {Znajek}.
\newblock {Electromagnetic extraction of energy from Kerr black holes}.
\newblock {\em \mnras}, 179:433--456, May 1977.

\bibitem{2017ApJ...848L..34M}
A.~{Murguia-Berthier}, E.~{Ramirez-Ruiz}, C.~D. {Kilpatrick}, et~al.
\newblock {A Neutron Star Binary Merger Model for GW170817/GRB 170817A/SSS17a}.
\newblock {\em \apjl}, 848:L34, October 2017.

\bibitem{2003MNRAS.343L..36R}
S.~{Rosswog} and E.~{Ramirez-Ruiz}.
\newblock {On the diversity of short gamma-ray bursts}.
\newblock {\em \mnras}, 343:L36--L40, August 2003.

\bibitem{2002MNRAS.336L...7R}
S.~{Rosswog} and E.~{Ramirez-Ruiz}.
\newblock {Jets, winds and bursts from coalescing neutron stars}.
\newblock {\em \mnras}, 336:L7--L11, October 2002.

\bibitem{2014ApJ...784L..28N}
H.~{Nagakura}, K.~{Hotokezaka}, Y.~{Sekiguchi}, M.~{Shibata}, and K.~{Ioka}.
\newblock {Jet Collimation in the Ejecta of Double Neutron Star Mergers: A New
  Canonical Picture of Short Gamma-Ray Bursts}.
\newblock {\em \apjl}, 784:L28, April 2014.

\bibitem{2005ApJ...625L..91R}
E.~{Ramirez-Ruiz}, J.~{Granot}, C.~{Kouveliotou}, et~al.
\newblock {An Off-Axis Model of GRB 031203}.
\newblock {\em \apjl}, 625:L91--L94, June 2005.

\bibitem{2014ApJ...788L...8M}
A.~{Murguia-Berthier}, G.~{Montes}, E.~{Ramirez-Ruiz}, F.~{De Colle}, and W.~H.
  {Lee}.
\newblock {Necessary Conditions for Short Gamma-Ray Burst Production in Binary
  Neutron Star Mergers}.
\newblock {\em \apjl}, 788:L8, June 2014.

\bibitem{2016ApJ...831...22F}
N.~{Fraija}, W.~H. {Lee}, P.~{Veres}, and R.~{Barniol Duran}.
\newblock {Modeling the Early Afterglow in the Short and Hard GRB 090510}.
\newblock {\em \apj}, 831:22, November 2016.

\bibitem{pop99}
R.~{Popham}, S.~E. {Woosley}, and C.~{Fryer}.
\newblock {Hyperaccreting Black Holes and Gamma-Ray Bursts}.
\newblock {\em \apj}, 518:356--374, June 1999.

\bibitem{nar01}
R.~{Narayan}, T.~{Piran}, and P.~{Kumar}.
\newblock {Accretion Models of Gamma-Ray Bursts}.
\newblock {\em \apj}, 557:949--957, August 2001.

\bibitem{rosIII}
Stephan {Rosswog}, Enrico {Ramirez-Ruiz}, and Melvyn~B. {Davies}.
\newblock {High-resolution calculations of merging neutron stars - III.
  Gamma-ray bursts}.
\newblock {\em \mnras}, 345(4):1077--1090, Nov 2003.

\bibitem{des08}
Luc Dessart, CD~Ott, Adam Burrows, Stefan Rosswog, and Eli Livne.
\newblock Neutrino signatures and the neutrino-driven wind in binary neutron
  star mergers.
\newblock {\em The Astrophysical Journal}, 690(2):1681, 2008.

\bibitem{perego2014neutrino}
Albino Perego, Stephan Rosswog, Ruben~M Cabez{\'o}n, et~al.
\newblock Neutrino-driven winds from neutron star merger remnants.
\newblock {\em Monthly Notices of the Royal Astronomical Society},
  443(4):3134--3156, 2014.

\bibitem{siegel2014magnetically}
Daniel~M Siegel, Riccardo Ciolfi, and Luciano Rezzolla.
\newblock Magnetically driven winds from differentially rotating neutron stars
  and x-ray afterglows of short gamma-ray bursts.
\newblock {\em The Astrophysical Journal Letters}, 785(1):L6, 2014.

\bibitem{gra05}
Jonathan Granot, Enrico Ramirez-Ruiz, and Rosalba Perna.
\newblock Afterglow observations shed new light on the nature of x-ray flashes.
\newblock {\em The Astrophysical Journal}, 630(2):1003, 2005.

\bibitem{ram05}
Enrico Ramirez-Ruiz, Jonathan Granot, Chryssa Kouveliotou, et~al.
\newblock An off-axis model of grb 031203.
\newblock {\em The Astrophysical Journal Letters}, 625(2):L91, 2005.

\bibitem{2019ApJ...871..123F}
N.~{Fraija}, F.~{De Colle}, P.~{Veres}, et~al.
\newblock {The Short GRB 170817A: Modeling the Off-axis Emission and
  Implications on the Ejecta Magnetization}.
\newblock {\em \apj}, 871:123, January 2019.

\bibitem{2019ApJ...871..200F}
N.~{Fraija}, A.~C.~C.~d.~E.~S. {Pedreira}, and P.~{Veres}.
\newblock {Light Curves of a Shock-breakout Material and a Relativistic
  Off-axis Jet from a Binary Neutron Star System}.
\newblock {\em \apj}, 871:200, February 2019.

\bibitem{2019arXiv190407732F}
N.~{Fraija}, D.~{Lopez-Camara}, A.~C. Caligula do E.~S. {Pedreira}, et~al.
\newblock {Signatures from a Cocoon and an off-axis material ejected in a
  merger of compact objects: An analytical approach}.
\newblock {\em arXiv e-prints}, page arXiv:1904.07732, Apr 2019.

\bibitem{cav78}
G.~{Cavallo} and M.~J. {Rees}.
\newblock {A qualitative study of cosmic fireballs and gamma-ray bursts}.
\newblock {\em \mnras}, 183:359--365, May 1978.

\bibitem{PIRAN1999575}
Tsvi Piran.
\newblock Gamma-ray bursts and the fireball model.
\newblock {\em Physics Reports}, 314(6):575 -- 667, 1999.

\bibitem{2004ApJ...608L...5L}
W.~H. {Lee}, E.~{Ramirez-Ruiz}, and D.~{Page}.
\newblock {Opaque or Transparent? A Link between Neutrino Optical Depths and
  the Characteristic Duration of Short Gamma-Ray Bursts}.
\newblock {\em \apjl}, 608:L5--L8, June 2004.

\bibitem{1994ApJ...430L..93R}
M.~J. {Rees} and P.~{Meszaros}.
\newblock {Unsteady outflow models for cosmological gamma-ray bursts}.
\newblock {\em \apjl}, 430:L93--L96, August 1994.

\bibitem{2017ApJ...848...15F}
N.~{Fraija}, P.~{Veres}, B.~B. {Zhang}, et~al.
\newblock {Theoretical Description of GRB 160625B with Wind-to-ISM Transition
  and Implications for a Magnetized Outflow}.
\newblock {\em \apj}, 848:15, October 2017.

\bibitem{1997ApJ...476..232M}
P.~{M{\'e}sz{\'a}ros} and M.~J. {Rees}.
\newblock {Optical and Long-Wavelength Afterglow from Gamma-Ray Bursts}.
\newblock {\em \apj}, 476:232--237, February 1997.

\bibitem{2015ApJ...804..105F}
N.~{Fraija}.
\newblock {GRB 110731A: Early Afterglow in Stellar Wind Powered By a Magnetized
  Outflow}.
\newblock {\em \apj}, 804:105, May 2015.

\bibitem{2016ApJ...818..190F}
N.~{Fraija}, W.~{Lee}, and P.~{Veres}.
\newblock {Modeling the Early Multiwavelength Emission in GRB130427A}.
\newblock {\em \apj}, 818:190, February 2016.

\bibitem{2017ApJ...848...94F}
N.~{Fraija}, W.~H. {Lee}, M.~{Araya}, et~al.
\newblock {Modeling the High-energy Emission in GRB 110721A and Implications on
  the Early Multiwavelength and Polarimetric Observations}.
\newblock {\em \apj}, 848:94, October 2017.

\bibitem{2019ApJ...885...29F}
N.~{Fraija}, S.~{Dichiara}, A.~C. Caligula do E.~S. {Pedreira}, et~al.
\newblock {Modeling the Observations of GRB 180720B: from Radio to Sub-TeV
  Gamma-Rays}.
\newblock {\em \apj}, 885(1):29, November 2019.

\bibitem{2019ApJ...883..162F}
N.~{Fraija}, R.~{Barniol Duran}, S.~{Dichiara}, and P.~{Beniamini}.
\newblock {Synchrotron Self-Compton as a Likely Mechanism of Photons beyond the
  Synchrotron Limit in GRB 190114C}.
\newblock {\em \apj}, 883(2):162, October 2019.

\bibitem{2019ApJ...879L..26F}
N.~{Fraija}, S.~{Dichiara}, A.~C. Caligula do E.~S. {Pedreira}, et~al.
\newblock {Analysis and Modeling of the Multi-wavelength Observations of the
  Luminous GRB 190114C}.
\newblock {\em \apjl}, 879(2):L26, July 2019.

\bibitem{wol78}
L.~{Wolfenstein}.
\newblock {Neutrino oscillations in matter}.
\newblock {\em \prd}, 17:2369--2374, May 1978.

\bibitem{2005MNRAS.364..934K}
H.~B.~J. {Koers} and R.~A.~M.~J. {Wijers}.
\newblock {The effect of neutrinos on the initial fireballs in gamma-ray
  bursts}.
\newblock {\em \mnras}, 364:934--942, December 2005.

\bibitem{1988NuPhB.307..924N}
D.~{N{\"o}tzold} and G.~{Raffelt}.
\newblock {Neutrino dispersion at finite temperature and density}.
\newblock {\em Nuclear Physics B}, 307:924--936, October 1988.

\bibitem{1991NuPhB.349..754E}
K.~{Enqvist}, K.~{Kainulainen}, and J.~{Maalampi}.
\newblock {Refraction and oscillations of neutrinos in the early universe}.
\newblock {\em Nuclear Physics B}, 349:754--790, February 1991.

\bibitem{1951PhRv...82..664S}
J.~{Schwinger}.
\newblock {On Gauge Invariance and Vacuum Polarization}.
\newblock {\em Physical Review}, 82:664--679, June 1951.

\bibitem{2014ApJ...787..140F}
N.~{Fraija}.
\newblock {Propagation and Neutrino Oscillations in the Base of a Highly
  Magnetized Gamma-Ray Burst Fireball Flow}.
\newblock {\em \apj}, 787:140, June 2014.

\bibitem{2004PhRvD..70d3001B}
E.~{Babaev}.
\newblock {Andreev-Bashkin effect and knot solitons in an interacting mixture
  of a charged and a neutral superfluid with possible relevance for neutron
  stars}.
\newblock {\em \prd}, 70(4):043001, August 2004.

\bibitem{1998PhRvD..58h5016E}
A.~{Erdas}, C.~W. {Kim}, and T.~H. {Lee}.
\newblock {Neutrino self-energy and dispersion in a medium with a magnetic
  field}.
\newblock {\em \prd}, 58(8):085016, October 1998.

\bibitem{2009PhRvD..80c3009S}
S.~{Sahu}, N.~{Fraija}, and Y.-Y. {Keum}.
\newblock {Neutrino oscillation in a magnetized gamma-ray burst fireball}.
\newblock {\em \prd}, 80(3):033009, August 2009.

\bibitem{2009JCAP...11..024S}
S.~{Sahu}, N.~{Fraija}, and Y.-Y. {Keum}.
\newblock {Propagation of neutrinos through magnetized gamma-ray burst
  fireball}.
\newblock {\em \jcap}, 11:24, November 2009.

\bibitem{1992PhRvD..46.1172D}
J.~C. {D'olivo}, J.~F. {Nieves}, and M.~{Torres}.
\newblock {Finite-temperature corrections to the effective potential of
  neutrinos in a medium}.
\newblock {\em \prd}, 46:1172--1179, August 1992.

\bibitem{2014MNRAS.437.2187F}
N.~{Fraija}.
\newblock {GeV-PeV neutrino production and oscillation in hidden jets from
  gamma-ray bursts}.
\newblock {\em \mnras}, 437:2187--2200, January 2014.

\bibitem{2015MNRAS.450.2784F}
N.~{Fraija}.
\newblock {Resonant oscillations of GeV-TeV neutrinos in internal shocks from
  gamma-ray burst jets inside stars}.
\newblock {\em \mnras}, 450:2784--2798, July 2015.

\bibitem{gon03}
M.~C. {Gonzalez-Garcia} and Y.~{Nir}.
\newblock {Neutrino masses and mixing: evidence and implications}.
\newblock {\em Reviews of Modern Physics}, 75:345--402, March 2003.

\bibitem{jarlskog1985c}
C~Jarlskog.
\newblock C. jarlskog, phys. rev. lett. 55, 1039 (1985).
\newblock {\em Phys. Rev. Lett.}, 55:1039, 1985.

\bibitem{PhysRevD.17.2369}
L.~Wolfenstein.
\newblock Neutrino oscillations in matter.
\newblock {\em Phys. Rev. D}, 17:2369--2374, May 1978.

\bibitem{mikheyev1986yad}
SP~Mikheyev.
\newblock Yad. fiz, 42 (1985) 1441; sp mikheyev, a. smirnov.
\newblock {\em Sov. J. Nucl. Phys}, 42:913, 1986.

\bibitem{RevModPhys.75.345}
M.~C. Gonzalez-Garcia and Yosef Nir.
\newblock Neutrino masses and mixing: evidence and implications.
\newblock {\em Rev. Mod. Phys.}, 75:345--402, Mar 2003.

\bibitem{aha11}
B.~{Aharmim} and et~al.
\newblock {Combined Analysis of all Three Phases of Solar Neutrino Data from
  the Sudbury Neutrino Observatory}.
\newblock {\em ArXiv e-prints}, September 2011.

\bibitem{abe11a}
K.~{Abe} and et~al.
\newblock {Search for Differences in Oscillation Parameters for Atmospheric
  Neutrinos and Antineutrinos at Super-Kamiokande}.
\newblock {\em Physical Review Letters}, 107(24):241801, December 2011.

\bibitem{chu02}
E.~D. {Church}, K.~{Eitel}, G.~B. {Mills}, and M.~{Steidl}.
\newblock {Statistical analysis of different $\nu^-_\mu$ -> $\nu^-_e$
  searches}.
\newblock {\em \prd}, 66(1):013001, June 2002.

\bibitem{1998PhRvL..81.1774}
C.~{Athanassopoulos} and et~al.
\newblock {Results on $\nu_\mu$ -> $\nu_e$ Neutrino Oscillations from the LSND
  Experiment}.
\newblock {\em Physical Review Letters}, 81:1774--1777, August 1998.

\bibitem{1996PhRvL..77.3082A}
C.~{Athanassopoulos} and et~al.
\newblock {Evidence for $\nu^-_\mu$ -> $\nu^-_e$ Oscillations from the LSND
  Experiment at the Los Alamos Meson Physics Facility}.
\newblock {\em Physical Review Letters}, 77:3082--3085, October 1996.

\bibitem{wen10}
R.~{Wendell} and et~al.
\newblock {Atmospheric neutrino oscillation analysis with subleading effects in
  Super-Kamiokande I, II, and III}.
\newblock {\em \prd}, 81(9):092004, May 2010.

\bibitem{salas2018}
PF~de~Salas, DV~Forero, CA~Ternes, M~T{\'o}rtola, and JWF Valle.
\newblock Status of neutrino oscillations 2018: 3$\sigma$ hint for normal mass
  ordering and improved cp sensitivity.
\newblock {\em Physics Letters B}, 2018.

\bibitem{dic72}
D.~A. {Dicus}.
\newblock {Stellar Energy-Loss Rates in a Convergent Theory of Weak and
  Electromagnetic Interactions}.
\newblock {\em \prd}, 6:941--949, August 1972.

\bibitem{lat76}
J.~M. {Lattimer}, C.~J. {Pethick}, M.~{Prakash}, and P.~{Haensel}.
\newblock {Direct URCA process in neutron stars}.
\newblock {\em Physical Review Letters}, 66:2701--2704, May 1991.

\bibitem{lun01}
Cecilia Lunardini and A~Yu Smirnov.
\newblock Supernova neutrinos: Earth matter effects and neutrino mass spectrum.
\newblock {\em Nuclear Physics B}, 616(1-2):307--348, 2001.

\bibitem{rom15}
Jorge Romao et~al.
\newblock {\em Neutrinos in high energy and astroparticle physics, 62,}.
\newblock John Wiley \& Sons, 2015.

\bibitem{smi05}
A~Yu Smirnov.
\newblock The msw effect and matter effects in neutrino oscillations.
\newblock {\em Physica Scripta}, 2005(T121):57, 2005.

\bibitem{kne08}
James~P. Kneller, Gail~C. McLaughlin, and Justin Brockman.
\newblock Oscillation effects and time variation of the supernova neutrino
  signal.
\newblock {\em Phys. Rev. D}, 77:045023, Feb 2008.

\bibitem{mur17}
A.~{Murguia-Berthier}, E.~{Ramirez-Ruiz}, G.~{Montes}, et~al.
\newblock {The Properties of Short Gamma-Ray Burst Jets Triggered by Neutron
  Star Mergers}.
\newblock {\em \apjl}, 835:L34, February 2017.

\bibitem{ree98}
M.~J. {Rees} and P.~{M{\'e}sz{\'a}ros}.
\newblock {Refreshed Shocks and Afterglow Longevity in Gamma-Ray Bursts}.
\newblock {\em \apjl}, 496(1):L1--L4, March 1998.

\bibitem{sar00}
Re'em {Sari} and Peter {M{\'e}sz{\'a}ros}.
\newblock {Impulsive and Varying Injection in Gamma-Ray Burst Afterglows}.
\newblock {\em \apjl}, 535(1):L33--L37, May 2000.

\bibitem{zha02}
Bing Zhang and Peter Meszaros.
\newblock Gamma-ray bursts with continuous energy injection and their afterglow
  signature.
\newblock {\em The Astrophysical Journal}, 566(2):712, 2002.

\bibitem{dai04}
ZG~Dai.
\newblock Relativistic wind bubbles and afterglow signatures.
\newblock {\em The Astrophysical Journal}, 606(2):1000, 2004.

\bibitem{rosII}
Stephan Rosswog and M~Liebend{\"o}rfer.
\newblock High-resolution calculations of merging neutron stars—ii. neutrino
  emission.
\newblock {\em Monthly Notices of the Royal Astronomical Society},
  342(3):673--689, 2003.

\bibitem{tub75}
D.~L. {Tubbs} and D.~N. {Schramm}.
\newblock {Neutrino Opacities at High Temperatures and Densities}.
\newblock {\em \apj}, 201:467--488, October 1975.

\bibitem{fuk03}
S~Fukuda, Y~Fukuda, T~Hayakawa, et~al.
\newblock The super-kamiokande detector.
\newblock {\em Nuclear Instruments and Methods in Physics Research Section A:
  Accelerators, Spectrometers, Detectors and Associated Equipment},
  501(2-3):418--462, 2003.

\bibitem{2011arXiv1109.3262A}
K.~{Abe} and et~al.
\newblock {Letter of Intent: The Hyper-Kamiokande Experiment --- Detector
  Design and Physics Potential ---}.
\newblock {\em ArXiv e-prints}, September 2011.

\bibitem{hk14}
{Hyper-Kamiokande Working Group}, {:}, K.~{Abe}, et~al.
\newblock {A Long Baseline Neutrino Oscillation Experiment Using J-PARC
  Neutrino Beam and Hyper-Kamiokande}.
\newblock {\em ArXiv e-prints}, December 2014.

\bibitem{hk18}
Ke~Abe, Ke~Abe, H~Aihara, et~al.
\newblock Hyper-kamiokande design report.
\newblock {\em arXiv preprint arXiv:1805.04163}, 2018.

\bibitem{acc16}
R~Acciarri, MA~Acero, M~Adamowski, et~al.
\newblock Long-baseline neutrino facility (lbnf) and deep underground neutrino
  experiment (dune) conceptual design report, volume 4 the dune detectors at
  lbnf.
\newblock {\em arXiv preprint arXiv:1601.02984}, 2016.

\bibitem{1989neas.book.....B}
J.~N. {Bahcall}.
\newblock {\em {Neutrino astrophysics}}.
\newblock 1989.

\bibitem{2004mnpa.book.....M}
R.~N. {Mohapatra} and P.~B. {Pal}.
\newblock {\em {Massive neutrinos in physics and astrophysics}}.
\newblock 2004.

\bibitem{2014MNRAS.442..239F}
N.~{Fraija}, C.~G. {Bernal}, and A.~M. {Hidalgo-Gam{\'e}z}.
\newblock {Signatures of neutrino cooling in the SN1987A scenario}.
\newblock {\em \mnras}, 442:239--250, July 2014.

\bibitem{dav14}
P~D'Avanzo, R~Salvaterra, MG~Bernardini, et~al.
\newblock A complete sample of bright swift short gamma-ray bursts.
\newblock {\em Monthly Notices of the Royal Astronomical Society},
  442(3):2342--2356, 2014.

\bibitem{jin18}
Zhi-Ping Jin, Xiang Li, Hao Wang, et~al.
\newblock Short {GRBs}: Opening angles, local neutron star merger rate, and
  off-axis events for {GRB}/{GW} association.
\newblock {\em The Astrophysical Journal}, 857(2):128, apr 2018.

\bibitem{patrignani2016review}
C~Patrignani, Particle~Data Group, et~al.
\newblock Review of particle physics.
\newblock {\em Chinese physics C}, 40(10):100001, 2016.

\end{thebibliography}
\addcontentsline{toc}{chapter}{Bibliography}\,.
\end{document}